\documentclass{aa}  

\usepackage{graphicx}
\usepackage{txfonts}
\usepackage{multirow}
\usepackage{multicol}
\usepackage{xcolor}
\usepackage{dirtytalk}
\usepackage{comment}
\usepackage[
raggedright
]{sidecap} 
\usepackage[switch]{lineno}
\usepackage{hyperref}
\usepackage{cleveref}
\crefformat{section}{\S#2#1#3} % see manual of cleveref, section 8.2.1
\crefformat{subsection}{\S#2#1#3}
\crefformat{subsubsection}{\S#2#1#3}
\begin{document}

 %  \title{Multi-scale, multi-frequency radio observations \\of the central AGN in the galaxy cluster RBS 797}

 \title{A JVLA, LOFAR, e-Merlin, VLBA and EVN study of RBS~797: deciphering the outburst history of the cluster-central radio galaxy}
 \title{A JVLA, LOFAR, e-Merlin, VLBA and EVN study of RBS~797: can binary SMBHs explain the outburst history of the central radio galaxy?}
\titlerunning{A JVLA, LOFAR, e-Merlin, VLBA and EVN study of RBS~797}
   \author{F. Ubertosi
          \inst{1,2}
           \and
           M. Giroletti
           \inst{3}
           \and
           M. Gitti
            \inst{1,3}
            \and
           N. Biava
           \inst{3,4}
           \and
           E. De Rubeis
           \inst{1,3}
           \and
           A. Bonafede
           \inst{1,3}
           \and
           L. Feretti
           \inst{3}
           \and
           M. Bondi
           \inst{3}
           \and
           L. Bruno
           \inst{1,3}
           \and
           E. Liuzzo
           \inst{3,5}
           \and
           A. Ignesti
           \inst{6}
           \and
           G. Brunetti
           \inst{3}
          %C. Ptolemy\inst{2}\fnmsep\thanks{Just to show the usage
          %of the elements in the author field}
          }

   \institute{$^{1}$ Dipartimento di Fisica e Astronomia, Università di Bologna, via Gobetti 93/2, I-40129 Bologna, Italy\\
    \email{francesco.ubertosi2@unibo.it} \\
    $^{2}$ Istituto Nazionale di Astrofisica - Osservatorio di Astrofisica e Scienza dello Spazio (OAS), via Gobetti 101, I-40129 Bologna, Italy\\
     $^{3}$ Istituto Nazionale di Astrofisica - Istituto di Radioastronomia (IRA), via Gobetti 101, I-40129 Bologna, Italy \\
     $^{4}$ Thüringer Landessternwarte, Sternwarte 5, 07778 Tautenburg, Germany \\
     $^{5}$ Italian Alma Regional Center (ARC), Via Piero Gobetti 101, I-40129 Bologna, Italy \\
     $^{6}$ INAF-Padova Astronomical Observatory, Vicolo dell’Osservatorio 5, 35122 Padova, Italy 
             }

   \date{Received 19 December 2023; Accepted 13 May 2024}

% \abstract{}{}{}{}{} 
% 5 {} token are mandatory
 
  \abstract
{}
   {The multi-faceted central radio galaxy of the cluster RBS~797 shows several episodes of jet activity in multiple directions. We wish to understand the causes behind these dramatic misalignments and measure the timescales of the successive outbursts.}
   {We present a multi-frequency (144~MHz -- 9~GHz) and multi-scale (5~pc -- $50$~kpc) investigation of the central radio galaxy in RBS~797, by means of JVLA, LOFAR (with international stations), e-Merlin, VLBA and EVN data. We investigate the morphological and spectral properties of the radio lobes, the jets, and the active core.}
   {We confirm the co-spatiality of the radio lobes with the four perpendicular X-ray cavities discussed in \citet{Ubertosi21apj}. The radiative ages of the E-W lobes ($31.4\pm6.6$~Myr) and of the N-S lobes ($32.1\pm9.9$~Myr) support a coeval origin of the perpendicular outbursts, that also have similar {\it active phase} duration ($\sim$12~Myr). Based on the analysis of the inner N-S jets (on scales of $\leq10$~kpc), we (a) confirm the S-shaped jet morphology; (b) show the presence of two hotspots per jet with a similar spectral index; (c) estimate the age of the twisting N-S jets to be less than $\sim8$~Myr. Based on these results, we determine that jet precession, with period $\sim$9~Myr, half-opening angle $\sim$24$^{\circ}$ and jet advance speed $\sim$0.01$c$, can explain the properties of the N-S jets. We also find that the synchrotron injection index has steepened from the large, older outbursts ($\Gamma\sim0.5$) to the younger S-shaped jets ($\Gamma\sim0.9$), possibly due to a transition from an FR~I-like to an FR~II-like activity. The e-Merlin, VLBA and EVN data reveal a single, compact core at the heart of RBS~797, surrounded by extended radio emission whose orientation depends on the spatial scale sampled by the data. }
   {We explore several engine-based scenarios to explain these results.  Piecing together the available evidence, we argue that RBS~797 likely hosts (or hosted) binary active SMBHs. The detection of a single component in the VLBI data is still consistent with this interpretation, since the predicted separation of the binary SMBHs ($\leq$0.6~pc) is an order of magnitude smaller than the resolution of the available radio data (5~pc).}


   \maketitle
%
%-------------------------------------------------------------------

\section{Introduction}
\label{sec:intro}
It is long known that supermassive black holes (SMBHs) can experience multiple phases of activity as Active Galactic Nuclei (AGN) over time (e.g., \citealt{saikia2009} and references therein). In the case of radio galaxies, the relativistic jets ejected from the central engine can extend from parsec (pc) to tens of kiloparsec (kpc) scales, where radio lobes are usually observed. After a few tens of Myr from the start of the jet activity, the central engine can turn off or greatly reduce its ability to collimate relativistic jets (e.g., \citealt{shabala2008,turner2015,morganti2017,hardcastle2020}). The fading radio lobes of sources in this phase usually show a steep radio spectrum and an amorphous shape \citep{murgia2011}. There are cases in which the SMBH drives a new pair of jets before the previous lobes completely fade. This can result in multiple pairs of radio lobes, with the oldest, most distant pairs having a steeper radio spectrum (e.g., \citealt{saikia2009,saikia2022,shabala2020}). 
\par A clear feature of AGN activity traced by restarted jets is the directionality of the successive episodes. 
A variety of radio sources have so far questioned the idea that SMBH-driven jets are always stable, straight, linear outflows. In the so-called \say{X-shaped} radio galaxies, two pairs of misaligned lobes are observed. Usually, these objects exhibit a pair of bright, primary jets, and a pair of fainter, secondary lobes (or \say{wings}) with a steeper spectral index (e.g., \citealt{leahy1992,bruno2019}). Additionally, even single pairs of jets can show signatures of wobbling. An S-shaped or Z-shaped morphology traces a curvature or sharp bending of the jet axis (e.g., \citealt{lister2003,Rubinur17}). In a few other cases, multiple active jets have been observed propagating from a galaxy's core (e.g., \citealt{hudson2006}). 
\par Several scenarios have been invoked to explain misaligned or wobbling jets. The wings of X-shaped radio galaxies may represent either an older episode of AGN activity in a different direction \citep{dennettthorpe2002,liu2004}, or backflow of the primary lobe plasma (e.g., \citealt{kraft2005}), or the jets from a secondary SMBH \citep{lal2007}. Binary SMBHs may also account for the observations of multiple pairs of jets ejected from a compact region (see e.g., the case of 3C~75, \citealt{hudson2006}). 
Ultimately, jets can experience geodetic precession over time (e.g., \citealt{fendt2022}). This mechanism is thought to be occurring in a large number of radio galaxies -- see e.g., \citealt{Krause19}, who found a fraction of $\sim$70\% of precessing jets in a complete sample of 3C radio galaxies.
Precession may naturally explain the S-shaped morphology of jets (see the cases of e.g., Hydra A \citealt{taylor1990}, Cygnus A, \citealt{Steenbrugge2008,Horton20}). 
\par Overall, it is still unclear if one of the above scenarios can univocally explain the signatures of the changes in jet axis. Combination of different mechanisms may be invoked. For instance, binary SMBHs can cause jet precession in the active SMBH due to the gravitational interaction of the binary (e.g., \citealt{begelman1980,Krause19}). 
In addition, a merger of binary SMBHs may lead to a spin-flip of the merged, more massive SMBH, which would drive jets in a perpendicular direction with respect to the previous outburst (e.g., \citealt{Merritt02}). The latter case would explain the jet reorientation events characterized by $\sim$90$^{\circ}$ misalignment between the successive jet activities. This holds especially for the radio galaxies with geometrically thick accretion disks, for which the Baarden-Petterson effect is inefficient (e.g., \citealt{fragile2007}).
\par A significant challenge in testing these scenarios arises from the vast range of spatial scales involved, spanning from a few pc to tens of kpc (see also \citealt{ubertosi23_cfa}). High-resolution and sensitive radio observations, that can target these widely varying spatial scales, are needed to understand how jets change their orientation over time. \\
\linebreak
The central radio galaxy of the galaxy cluster RBS~797 (RA 09:47:12.76, DEC +76:23:13.74, $z = 0.354$) represents an exemplary case of multiple and misaligned jet activities. 
In the innermost 50 kpc, multiple pairs of radio lobes that are misaligned by roughly 90$^{\circ}$ have been discovered \citep{Gitti06,Gitti13}, with the largest and brightest lobes extended in the east (E) - west (W) direction (at $\sim$30 kpc from the center), radio fainter lobes in the north (N) - south (S) direction (at $\sim$27 kpc from the center) and twin jets heading N-S in the inner $\sim$10 kpc. 
At sub-arcsec resolution, a short observation of the radio core with the European VLBI Network (EVN) at 5 GHz uncovered two compact components separated by $\sim$77 pc \citep{Gitti13}. Combining these findings, \citet{Gitti13} suggested that either the central SMBH has experienced reorientation events, producing subsequent outbursts in perpendicular directions, or that the brightest cluster galaxy (BCG) hosts a dual AGN, with the two active SMBHs simultaneously launching jets in perpendicular directions (coeval outbursts). 
\par
\begin{figure*}[ht!]
   \centering
   \sidecaption
   \includegraphics[width=0.7\linewidth]{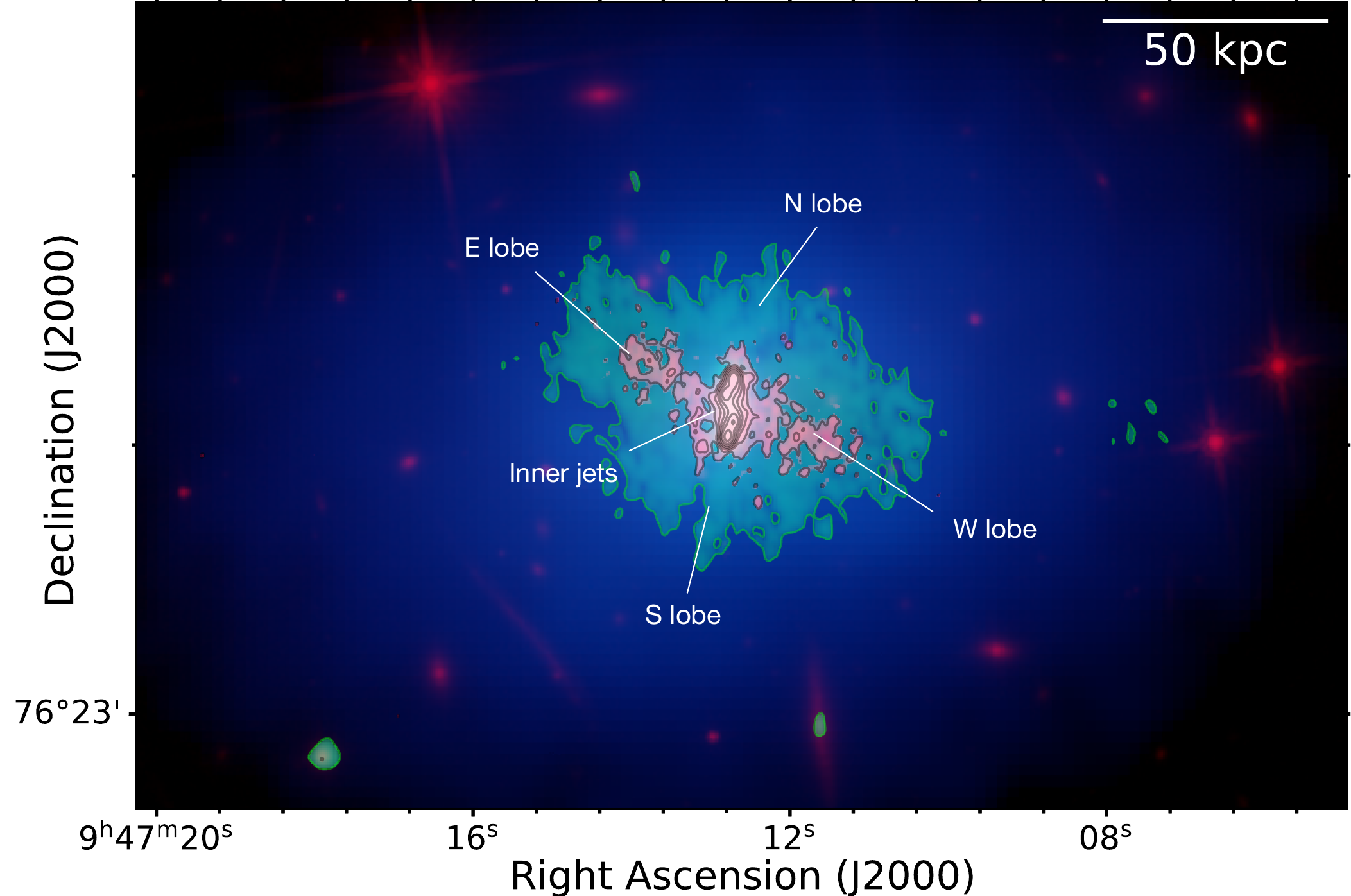}
      \caption[Composite optical (HST, red), $Chandra$ (X-ray, blue), 3 GHz (JVLA, green), 144 MHz (LOFAR-International, pink) image of RBS~797.]{Composite optical (HST, red), $Chandra$ (X-ray, blue), 3 GHz (JVLA, green), 144 MHz (LOFAR-International, pink) image of RBS~797. The structures discussed in this work are labelled. Black contours (described in Fig. \ref{fig:jvlalofar}) outline the 144 MHz emission of the radio galaxy.
              }
         \label{fig:composite}
   \end{figure*}

 Recently, deep {\it Chandra} data showed that the AGN activity left numerous imprints on the intracluster medium (ICM) of RBS~797 \citep{Ubertosi21apj,Ubertosi23apj}. The detection of strong depressions in the X-ray surface brightness revealed that two perpendicular and equidistant pairs of X-ray cavities were carved in the hot gas at $\sim$30 kpc from the center by the perpendicular and equidistant radio lobes \citep{Ubertosi21apj}. Additionally, three nested shock fronts driven by the AGN activity were found at distances of 50 kpc, 80 kpc and 130 kpc from the center. The inner shock has the shape of a whole cocoon which surrounds the four perpendicular and equidistant X-ray cavities \citep{Ubertosi23apj}. A striking result based on the X-ray data was that the two perpendicular pairs of X-ray cavities are nearly coeval, around 30 Myr old, with an age difference of $1\pm7$~Myr. \citet{Ubertosi21apj} concluded that this relatively short outburst interval may be explained either by a nearly coeval activity of active binary SMBHs, or by a rapid reorientation (a few Myr at most) of the jets from a single AGN. \\
\linebreak
\noindent In this work, we present new multi-frequency radio observations of the radio galaxy in RBS~797 from different instruments. We show in Fig. \ref{fig:composite} a composite image of RBS~797, created with our new LOFAR and JVLA data. We present the data and the reduction techniques in Section \cref{sec:data}. We show our results in Section \cref{sec:results}, by focusing on the kpc scale (\cref{subsec:kpc}), the kpc - pc connection (\cref{subsec:kpcpc}), and the pc scale (\cref{subsec:pc}). The results are discussed in Section \cref{sec:discussion}, and we summarize our work in Section \cref{sec:conclusion}.
\par Throughout this work we assume a $\Lambda$CDM cosmology with H$_{0}$=70~km~s$^{-1}$~Mpc$^{-1}$, $\Omega_{\text{m}}=0.3$, and $\Omega_{\Lambda}=0.7$, which gives a scale of 4.9 kpc~arcsec$^{-1}$ at $z=0.354$. We define the radio spectral index $\alpha$ with the following convention: $S_{\nu} \propto \nu^{-\alpha}$, where $S_{\nu}$ is the flux density at a frequency $\nu$. Unless otherwise stated, uncertainties are reported at 1$\sigma$.
%--------------------------------------------------------------------
\section{Radio observations of RBS~797: data reduction}
\label{sec:data}
\subsection{VLBA, EVN and e-Merlin data reduction}
\label{subsec:vlbi}
\label{obsdata}
The observations of the BCG in RBS 797 comprise new data from the Very Long Baseline Array (VLBA) 
%, \citealt{Beasley95}
using bands L, C and X; the European VLBI Network (EVN) using bands L and C; and the Multi-Element Radio-Linked Interferometer Network (eMerlin) using bands L and C. In the following subsections we describe the observations and the data reduction procedure for the different arrays and frequencies considered. A summary of the observations presented in this article is provided in Tab. \ref{tab:sumobs}. 

\subsubsection{VLBA and EVN data}
\label{subsubsec:vlbaevn}
The radio core of RBS~797 has been observed by the VLBA in L band (1.6 GHz), C band (5 GHz) and X band (8.4 GHz) in phase-referencing mode (project code: BG224, PI: Gitti), with 8 spectral windows (each with a bandwidth of 32 MHz and 128 channels). For all the observing frequencies, the source J0954+7435 has been used as phase calibrator, while the source J2005+7752 has been used as fringe finder. The target has been observed in 4 min scans bracketed by 60 s scans of the phase calibrator, for a total time-on-source of 4 h in L band (60 scans on the target), 40 min in C band (10 scans on the target), and 1 h 20 min in X band (20 scans on the target).
\linebreak
\\The BCG of RBS~797 has been targeted with the EVN in phase-referencing mode (with 8 spectral windows, each with a bandwidth of 16 MHz and 32 channels). The BCG was observed at 1.6 GHz (project code: EG080-A, PI: Gitti) and at 5 GHz (project code: EG080-B, PI: Gitti), respectively. At both frequencies, J0954+7435 has been used as phase calibrator (observed in scans of 60 s), while 3C345 was observed as fringe finder. The target has been observed in 3m 30s scans for a total time-on-source of $\sim$5h at each frequency.
\linebreak
\\The strategy adopted to process the data in AIPS \citep{Greisen03} is similar for the VLBA and EVN observations. We followed standard reduction techniques, namely:
\begin{itemize}
	\item Correction of Earth orientation parameters and ionospheric delays (tasks \texttt{VLBAEOPS, VLBATECR}). Removal of bad data from system temperatures of the EVN data.
	\item Correction of delays and amplitudes; we applied digital sampling corrections and removed instrumental delays on phases (\texttt{VLBACCOR} \texttt{VLBAMPCL} for the VLBA, \texttt{FRING} for the EVN). Then, we calibrated the bandpass (\texttt{BPASS}). At this point, we removed bad data from system temperatures and gain curves of the VLBA observations. To complete the amplitude calibration, we applied the corrections with the task \texttt{VLBAAMP}.
	\item The time-dependent delays, rates, and phases where corrected by fringe-fitting the data of the calibrator (\texttt{FRING}). 
	To properly correct the visibilities we used an image of the calibrator as a model for fringe-fitting. 
	\item Ultimately, we applied the calibration to the data and averaged the different spectral windows. After editing the final visibilities, self-calibration was attempted only for the L-band data (due to the low flux density of the target), finding however no significant improvement in dynamical range or sensitivity. Images were made in AIPS with the task \texttt{IMAGR}.
\end{itemize}
We found that supplying an image model during fringe-fitting is crucial when J0954+7435 is used as a calibrator: since the calibrator itself is resolved into a double component, assuming a point source as a model for fringe-fitting results in a bad description of rate and delays (see also \cref{subsec:pc}). 
\par When using the software \texttt{DIFMAP} \citep{Shepherd97} to fit visibilities with gaussian-like or point-like components, the uncertainties on flux densities were computed as: 
\begin{equation}
    \delta S_{\nu} = \sqrt{N_{beam}\sigma_{\text{rms}}^{2} + (f_{\%}S_{\nu})^{2} + \delta_{difmap}^{2} }
\end{equation}
where $N_{beam}$ is the ratio between the area within which the flux density is computed and the beam area, $\sigma_{\text{rms}}$ is the local r.m.s. noise, $f_{\%}$ is the relative uncertainty on the flux density scale (here we assume $f_{\%} = 10\%$), $S_{\nu}$ is the flux density indicated by the fit and $\delta_{difmap}$ is the statistical fit uncertainty (at 1$\sigma$ confidence level, typically around 5\%$S_{\nu}$) returned by the fit in \texttt{DIFMAP}. 
\setlength{\tabcolsep}{3pt}
\begin{table*}
    \centering
    \caption[Summary of the EVN, VLBA, e-Merlin, JVLA and LOFAR data presented in this work.]{Summary of the EVN, VLBA, e-Merlin, JVLA and LOFAR data presented in this work.}
    \renewcommand{\arraystretch}{1.5}
\begin{tabular}{c|c|c|c|c|c}
\hline
 \multicolumn{6}{c}{{\large PRIMARY DATA}} \\
\hline
 (1) & (2)  & (3) & (4) & (5) & (6) \\
\hline
 Telescope & Project Code  & Obs. Date & Time on source & Frequency & Calibrators \\
 &  &  & [h] & [GHz] &  \\
 \hline

 \multirow{3}{*}{JVLA}    & \multirow{3}{*}{22A-301} & \multirow{3}{*}{May 2022} & 1.5 & 3.0 & \multirow{3}{*}{3C286, J1044+8054} \\
 
  &  &  & 0.7 & 5.5 &  \\
 
  &  &  & 2 & 9.0 & \\

 \hline

 LOFAR & LC~10\_010 & June 2018 & 8 & 0.144 & 3C295, L619462 \\
 
 \hline
 
 \multirow{2}{*}{eMerlin}  & \multirow{2}{*}{EG080} & March 28, 2014 & 6 & 1.6 & 0319+415, J0954+7435  \\
%6h (54$\times$7min)
  &  & March 10, 2014 & 6 & 5.0 & 1331+305, J0954+7435 \\
%{\color{red}27h?? (237$\times$7min)}
\hline
 
 \multirow{3}{*}{VLBA}  & \multirow{3}{*}{BG~224} & \multirow{3}{*}{February 4, 2014} & 4 & 1.6 & \multirow{3}{*}{J2005+7752, J0954+7435}  \\
 
  &  & & 0.7 & 5.0 &  \\
 
  &  &  & 1.3 & 8.4 &  \\

 \hline
 \multirow{2}{*}{EVN} & \multirow{2}{*}{EG080} & February 26, 2014 & 5 & 1.6 & \multirow{2}{*}{3C345, J0954+7435} \\

  &  & March 9, 2014 & 5 & 5.0 &  \\

\hline

 \multicolumn{6}{c}{{\large COMPLEMENTARY DATA}} \\
 
%\hline
% Telescope & Project Code  & Date & Time on source & Frequency & Calibrators & PI  \\
 \hline

  \multirow{2}{*}{EVN} & RSG05 & May 3, 2013 & 1 & 5.0 & 0133+476, J0954+7435 \\

  & RSG07 & December 1, 2015 & 1.5 & 1.6 & J0958+6533, J1027+7428, J0954+7435 \\
 \hline
  VLBA & BE~056 & July 8, 2014 & 0.3 & 5.0 & 4C~39.25, J1044+8054 \\
  \hline

\end{tabular}\label{tab:sumobs}
\tablefoot{We separate the primary, sensitive data (that we use to derive the main results) from the shorter-observing time complementary data (that we use to verify the results based on primary data). (1) Name of the radio telescope; (2) project code; (3) date of the observations; (4) time spent on the target source RBS~797; (5) observing frequency; (6) Name of the calibrators.}
\end{table*}

\begin{table*}
    \centering
    \caption{Image parameters of the EVN, VLBA, e-Merlin, JVLA and LOFAR data used in this paper.}
    \renewcommand{\arraystretch}{1.5}
\begin{tabular}{c|c|c|c|c|c|c|c|c|c}
\hline
 (1) & (2)  & (3) & (4) & (5) & (6) & (7) & (8) & (9) & (10)\\
\hline
Telescope & Project Code  & Frequency & Robust & Beam & P.A. & $\sigma_{\text{rms}}$ & Peak & $S_{\nu}$ ($5\sigma_{\text{rms}}$) &  Figure \\
&  & [GHz] &  & [mas] & [$^{\circ}$] & [$\mu$Jy/beam] & [mJy/beam] & [mJy] &   \\
 \hline

 \multirow{4}{*}{\color{black}JVLA}    & \multirow{4}{*}{22A-301}   &       3.0 & \texttt{2} & $900\times900$ & 0& 5.0 & 2.07  & 10.09 $\pm$ 0.50  &  Fig. \ref{fig:jvlalofar}\textit{a}\\

& & $3.0$ & \texttt{0} & $590\times400$ & 5.7& 6.0 & 1.69 & 4.89 $\pm$ 0.24 &  Fig. \ref{fig:jvlalofar}\textit{b}\\

    &  &       5.5 & \texttt{0} & $388\times250$ & 15.4 & 5.5 & 1.57  & 2.97 $\pm$ 0.15 & Fig. \ref{fig:jvlalofar}\textit{c}\\
 
    & &       9.0 & \texttt{0} & $230\times200$ & -4.5 & 3.0 & 1.36  & 2.07 $\pm$ 0.10 &  Fig. \ref{fig:jvlalofar}\textit{d}\\

\hline
{\color{black}LOFAR }   & LC 10\_010   &   0.144 & \texttt{0} & $360\times250$ & 170.4 & 40  & 13.0  & 134.4 $\pm$ 26.9  & Fig. \ref{fig:jvlalofar}\\
\hline

\multirow{2}{*}{\color{black}eMerlin} & \multirow{2}{*}{EG080}   &      1.6 & \texttt{5} & $150\times150$ & 0& 14  & 1.51 & 3.44 $\pm$ 0.20 & \multirow{2}{*}{Fig. \ref{fig:merlin}} \\

 &   &      5.0 & \texttt{0} & $35\times35$ & 0& 20  & 0.50  & 0.94 $\pm$ 0.08 &\\

\hline

 \multirow{4}{*}{\color{black}VLBA}   &  \multirow{3}{*}{BG~224} &      1.6 & \texttt{0} & $6.3\times4.6$ & 82.3 & 25  & 0.61  & 0.94 $\pm$ 0.09  & \multirow{3}{*}{Fig. \ref{fig:vlbaevn}}\\

   &     &      5.0 & \texttt{0} & $2.2\times1.4$ & -80.9 & 30  & 0.38  & 0.47 $\pm$ 0.05  &\\
 
   &     &      8.4 & \texttt{0} & $1.3\times0.9$ & -86.4 & 30  & 0.39  & 0.40 $\pm$ 0.05 & \\
 
    \cline{2-10}
    &  BE~056 & 5.0 &  \texttt{3} & $5.6\times4.6$ & 7.17 & 25  & 0.35  & 0.47 $\pm$ 0.06  & -- \\
 
\hline

\multirow{4}{*}{\color{black}EVN}     & \multirow{2}{*}{EG080}   &      1.6 & \texttt{0} & $9.1\times6.7$ & -46.9 & 30  & 1.37  & 1.65 $\pm$ 0.17  & \multirow{2}{*}{Fig. \ref{fig:vlbaevn}}\\

     &    &      5.0 & \texttt{0} & $2.6\times1.6$ & -62.6 & 15  & 0.51  & 0.70 $\pm$ 0.07 &\\

\cline{2-10}
     & RSG05     &      5.0 & \texttt{0} & $7.2\times5.2$ & -13.9 & 20  & 0.68   & 0.84 $\pm$ 0.09  & --\\
\cline{2-10}
     & RSG07     &      1.6  & \texttt{0} & $29.1\times23.9$ & -36.4 & 64  & 1.50  & 1.91 $\pm$ 0.22  & --\\

\hline

\end{tabular}\label{tab:imagepar}
\tablefoot{(1) Name of the radio telescope; (2) project code; (3) observing frequency; (4) \texttt{robust} parameter used during imaging; (5) beam major and minor axes; (6) beam position angle; (7) r.m.s. noise; (8) peak flux of the image; (9) total flux density within the $5\sigma_{\text{rms}}$ contour; (10) corresponding figure. The weighting obtained with a specific absolute value of the \texttt{robust} parameter depends on the processing software: the parameter refers to the CASA task \texttt{tclean} for the JVLA data, to the WSClean software for the LOFAR data, and to the AIPS task \texttt{IMAGR} for the e-Merlin, VLBA and EVN data. The EVN project RSG05 had already been published in \citet{Gitti13}, and a preliminary JVLA image at 3.0~GHz from project 22A-301 (with resolution 0.9''$\times$0.8'') was presented in \citealt{Ubertosi23apj}.}
\end{table*}

\subsubsection{e-Merlin data}
\label{subsubsec:emerlin}
The e-Merlin observations of RBS 797 (project code EG080, PI: Gitti) have been performed at L band (1.6 GHz) and C band (5 GHz). 
At both frequencies, the source J0954+7435 has been used as phase calibrator. The fringe finders were 0319+415 for the L band data and 1331+305 for the C band data.
%Calibration by beswick:
After being passed through the e-MERLIN pipeline \citep{Argo15}, any problematic data such as spurious signals or data from faulty antennas were identified and flagged. 
\par Images were made in AIPS with the task \texttt{IMAGR}. We found that the best results are achieved by setting the \texttt{robust} parameter to \texttt{R} $=0$ for the C band data and by selecting a natural weighting for the L band data.  
Peak and total flux densities (above the $5\times\sigma_{\text{rms}}$ contour) of the radio emission were measured in AIPS at the different frequencies, with the uncertainties being computed as:
\begin{equation}
\label{eq:uncertflux}
    \delta S_{\nu} = \sqrt{N_{beam}\sigma_{\text{rms}}^{2} + (f_{\%}S_{\nu})^{2}},
\end{equation}
\noindent where we assume the typical flux density scale uncertainty of $f_{\%} = 5\%$.
By inspecting the radio images, we noticed that there is a shift in the astrometry of the e-Merlin observations, by $-38$ milliarcseconds in RA and $+3$ milliarcseconds in DEC. In order to properly compare the e-Merlin results with those obtained from other facilities, we corrected the data for this astrometric offset by shifting the images in AIPS.

\subsection{JVLA observations and data reduction}\label{subsec:jvla}
We analyze new {\it Karl Jansky} Very Large Array (JVLA) observations of RBS~797 (Project code 22A-301, PI: Ubertosi) performed with the A array at S band (2 -- 4 GHz), C band (4.5 -- 6.5 GHz) and X band (8 -- 10 GHz). In all observations the source 3C~286 (J0137+3309) was used as the primary flux density calibrator, while J1044+8054 was used as phase calibrator. The data were calibrated in CASA v.6.4.1 \citep{casateam2022} using standard data reduction techniques for continuum calibration\footnote{See \url{https://casaguides.nrao.edu/}.}. Self-calibration of the target is not possible given its relatively low flux density (peak flux density of $1-2$~mJy at GHz frequencies, see Tab. \ref{tab:imagepar}). Nonetheless, the noise in our images is close to the theoretical expected one.
\par
\begin{figure*}[ht!]
   \centering
   \includegraphics[width=\hsize]{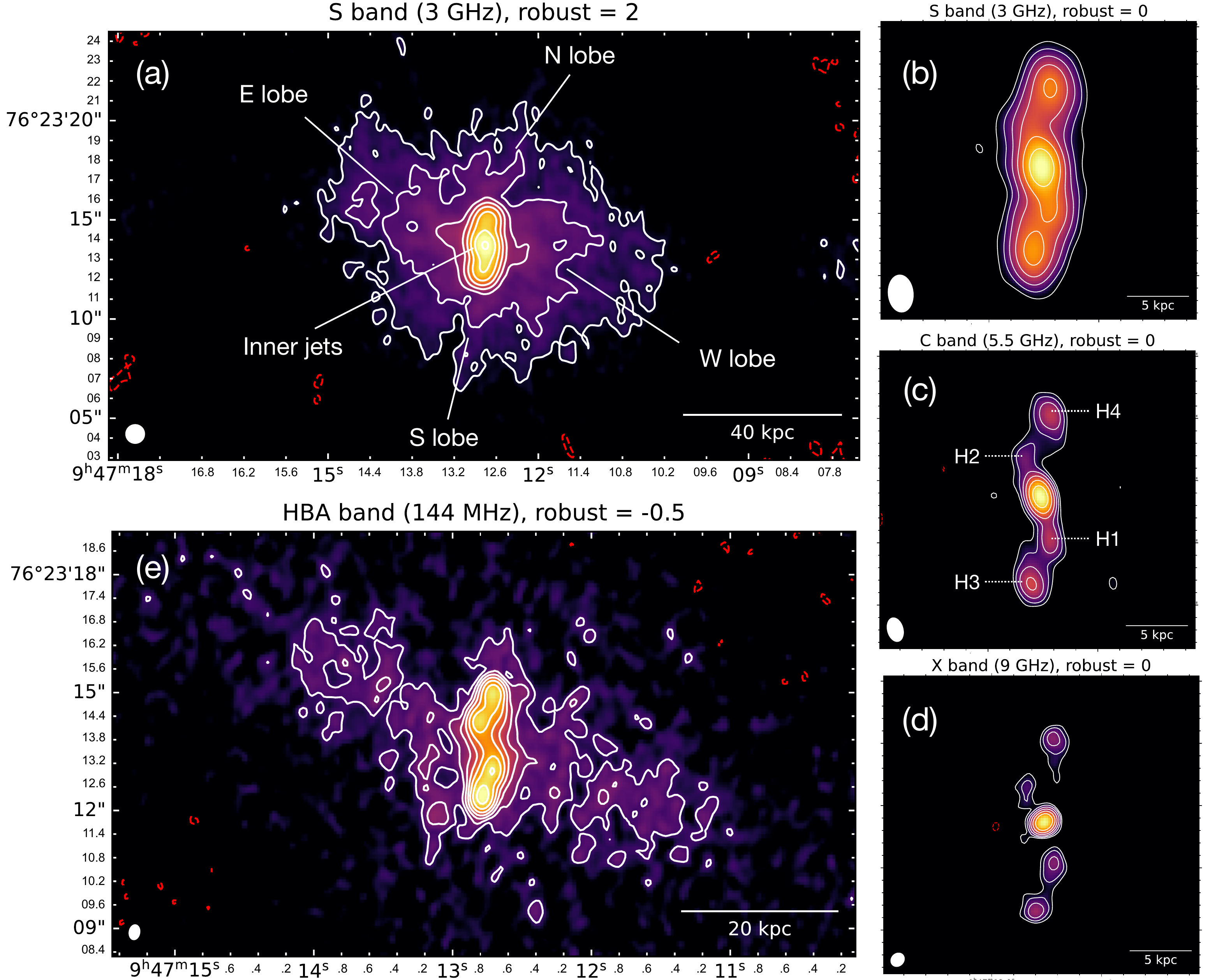}
      \caption{Final JVLA and LOFAR maps of RBS~797. (a) 3 GHz JVLA image at 0.9$''$ resolution, obtained by setting the \texttt{robust} parameter to 2. (b), (c), (d) 3 GHz, 5.5 GHz, and 9 GHz JVLA images obtained by setting \texttt{robust = 0}. The beam sizes are 0.59$''\times$0.40$''$ at 3~GHz, 0.39$''\times$0.25$''$ at 5.5~GHz, and 0.23$''\times$0.20$''$ at 9~GHz. (e) 144 MHz LOFAR image with the International Stations at 0.36$''\times$0.25$''$ resolution, obtained by setting \texttt{robust = -0.5}. In each panel, the beam is shown with a white circle in the bottom left corner. In panels (a) and (e), contours start at 3$\sigma_{\text{rms}}$ and increase by a factor of 2; in panels (b), (c), and (d), contours start at 5$\sigma_{\text{rms}}$ and increase by a factor of 2. The first negative contour ($-3\sigma_{\text{rms}}$ or $-5\sigma_{\text{rms}}$) is plotted in dashed red. See Tab. \ref{tab:imagepar} for the $\sigma_{\text{rms}}$ noise levels of the images.
      Structures discussed in this work are labelled.}
         \label{fig:jvlalofar}
   \end{figure*}

 The data were imaged in CASA using the task \texttt{tclean}, and testing different weighting combinations. For the S band (3.0 GHz) data, we find that an image made with \texttt{robust = 0} highlights the jet structure in the inner 15 kpc, while an image made with \texttt{robust = 2} recovers extended emission on larger scales, reaching the lobe of the radio galaxy. For the data at C and X bands, the images that best capture tha radio morphology are obtained by setting \texttt{robust = 0}. Different choices of weighting schemes were tested, but the image quality or the total recovered flux density did not demonstrate any significant variations. The images are shown in Fig. \ref{fig:jvlalofar} (panels \textit{a}, \textit{b}, \textit{c}, \textit{d}). 
We measured the peak and total flux density (above the $5\times\sigma_{\text{rms}}$ contour) of the radio emission at the different frequencies. The uncertainties were computed as in Eq. \ref{eq:uncertflux}, assuming the typical flux density scale uncertainty of 5\%.
\subsection{LOFAR long-baseline observations}\label{subsec:lofar}
RBS~797 has been observed by the LOw Frequency ARray (LOFAR, \citealt{vanHaarlem2013}) in 2018 (Project code: LC~10\_010, PI: Bonafede). The observations used the HBA antennas and the same setup of the LOFAR Two Meter Sky Survey (LoTSS, e.g., \citealt{Shimwell22}), with 22	core stations, 13 remote stations, and 13 international stations. The source has been observed between 120 MHz and 168 MHz (central frequency 144\,MHz) for 8h, using the calibrator 3C295.  
\par To calibrate the data (including the international stations) we followed the prescriptions detailed in \citet{Morabito22}\footnote{See also \url{https://lofar-vlbi.readthedocs.io}.} using the HOTCAT High Performance Computing cluster at INAF Trieste \citep{Taffoni20,Bertocco20}. The data were first passed through the \texttt{Pre-factor3} software package \citep{vanWeeren16,Williams16,deGasperin19}, which uses the calibrator source to derive the corrections for polarization alignment, Faraday rotation, bandpass, and clock offsets. Then, a first set of phase calibration solutions for the core and remote stations (derived using a sky model of the field from the TGSS survey, \citealt{Intema17}) are added to the calibrator solutions. 
\par The LOFAR-VLBI pipeline is subsequently used to calibrate the international stations. This requires to select a bright and compact source separated by less than 1 degree from the target from the Long-Baseline Calibrator Survey (LBCS, \citealt{Jackson22}). We derived the solutions for the international stations using L619462, which is located at 0.56$^{\circ}$ from the target RBS~797. After splitting out the target visibilities, the solutions from the delay calibrator were applied to the target. 
\par To improve the image quality, self-calibration was performed on the target using the Default Preprocessing Pipeline DPPP \citep{vanDiepen18} to find the solutions, and  WSClean v3.2 \citep{Offringa14} to obtain images (see also \citealt{vanweeren2021}). We performed three cycles of self-calibration solving for total electron content (TEC) and phase, and a final cycle of amplitude corrections. To correct for possible offset in the absolute flux density scale (see \citealt{Shimwell22}), we compared the flux density of the unresolved delay calibrator between the LOFAR-VLBI image and the 6$''$ LOFAR survey image, finding a correction factor of 1.61 that was applied to the images. We computed the uncertainties on the flux densities using Eq. \ref{eq:uncertflux} and assuming a 20\% uncertainty on the absolute flux density scale to account for uncertainties in the above correction factor of LOFAR-VLBI data (see also \citealt{Mahatma23}). This conservative choice is higher than the typical 10\% assumed for LOFAR data \citep{Shimwell22}. The angular resolution of the final image is of 0.36$''\times$0.25$''$, and the r.m.s. noise is $\sigma_{\text{rms}} = 40\,\mu$Jy/beam. Images with a resolution down to a $\sim$1$''$ beam were tested, but no improvement was found. The missing uv-coverage between baselines of 50k$\lambda$ -- 75k$\lambda$ \citep{Morabito22} prevents us from trading off high spatial resolution for improved image sensitivity (see also \cref{subsubsec:speclobes}). We point the reader to \citet{bonafede2023} for a study of diffuse emission in RBS~797 using sensitive, low-resolution (beam of $\sim$4$''$) LOFAR data.
\subsection{Complementary archival data}\label{subsec:archive}
To complement the analysis of the targeted, deep observations presented above we consider additional archival data from different facilities. We use two EVN snapshot observations, one at 1.6 GHz (project RSG07), and the other at 5 GHz (project RSG05, published in \citealt{Gitti13}). We also use a snapshot VLBA observation at 5 GHz to understand if time variability is present in RBS~797 (project BE056, PI Edge; see \cref{subsubsec:variab}). We reduced the data with the same approach presented in \cref{subsec:vlbi}. A summary of the observations is reported in Tab. \ref{tab:sumobs} and \ref{tab:imagepar}.
\par Additionally, we relied on archival VLA data at 1.4 GHz (arrays A, B, and C; see \citealt{Gitti06,Doria12}), and at 4.8 GHz (arrays A and B, see \citealt{Gitti13}). These data were used in combination with our new JVLA and LOFAR observations to measure the spectral properties of the radio galaxy (see \cref{subsubsec:speclobes}).
     \begin{figure*}[ht!]
   \centering
   \includegraphics[width=0.9\hsize]{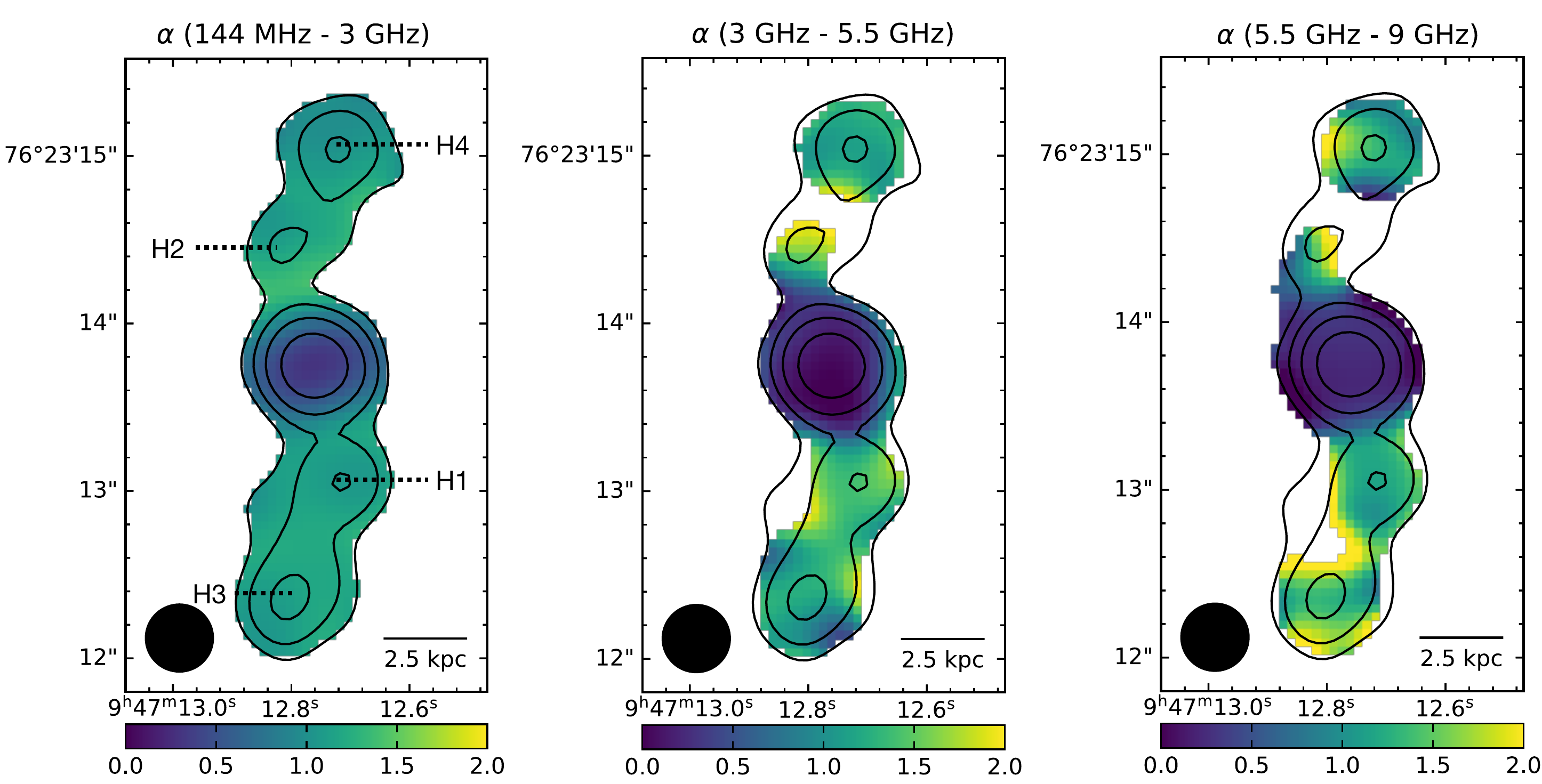}
      \caption[Spectral index maps of the inner jets]{
      %{\it Top panels}: LOFAR and JVLA maps used to compute the spectral index maps, at a resolution of 0.4''. The beam is shown in the bottom left corner. Contours start at 5$\sigma_{\text{rms}}$ and increase by a factor of 2, with the $\sigma_{\text{rms}}$ of the maps being 43 $\mu$Jy/beam (144 MHz), 22 $\mu$Jy/beam (3 GHz), 12 $\mu$Jy/beam (5.5 GHz), 7 $\mu$Jy/beam (9 GHz). 
      %{\it Bottom panels}: 
    Spectral index maps of the inner jets. The maps are between 144 MHz and 3 GHz (left), 3 GHz and 5.5 GHz (center), 5.5 and 9 GHz (right), displayed with a matching color-scale of $0\leq\alpha\leq2$. Contours from the 3 GHz map used for the spectral index are overlaid. The relative uncertainty on the spectral index range between 2.5\% - 15\% (144 MHz - 3 GHz), 6\% - 15\% (3 GHz - 5.5 GHz), 7\% -- 30\% (5.5 - 9 GHz). Structures discussed in this work are labelled.}
         \label{fig:jvlalofar-spix}
   \end{figure*}
\section{Results}\label{sec:results}
Given the large difference in physical scales to which the employed data are sensitive (from tens of kpc down to tens of pc), in this section we present our results by progressively zooming into the radio core of RBS~797. Starting from the kpc scales, we present the new observations performed with the JVLA and LOFAR telescopes. Then, we probe the connection between the kpc and pc scales using the e-Merlin observations. Ultimately, we present the images obtained from the EVN and VLBA data, that allow us to peer into the pc scales.
\subsection{The kpc scale}\label{subsec:kpc}
\subsubsection{Radio galaxy morphology}\label{subsubsec:radiogal}
We show in Fig. \ref{fig:jvlalofar} the images obtained from the JVLA data at 3 GHz (panels \textit{a} and \textit{b}), 5.5 GHz (panel \textit{c}), and 9 GHz (panel \textit{d}) and the image obtained from the LOFAR-VLBI data at 144 MHz (panel \textit{e}). The 3 GHz JVLA image obtained with \texttt{robust = 2} unveils the full morphology of the radio galaxy at a resolution of 0.9$''$ (Fig. \ref{fig:jvlalofar}\textit{a}; a preliminary image was already presented in \citealt{Ubertosi23apj}). This image allows one to appreciate the multi-faceted morphology of the radio galaxy in RBS~797, with its perpendicular pairs of radio lobes, extending for $\sim$50~kpc, that fill all the four X-ray cavities in the ICM (see also \citealt{Ubertosi21apj}), and the inner north-south jets of the AGN. 
\par Such north-south inner jets, with a distinctive S-shape, are better visible in the maps at 3.0 GHz ($\sim$0.5$''$ resolution), 5.5 GHz ($\sim$0.3$''$ resolution), and 9 GHz ($\sim$0.2$''$ resolution) shown in the right panels. The jets extend for $1.8''$ ($\sim$9 kpc) on each side of the core. We identify a total of four bright hotspots, two on each sides of the nucleus and symmetric with respect to the core. 
The connection between the inner jets and the larger, diffuse radio emission is clearer in the LOFAR image with international stations ($\sim$0.3$''$ resolution). At 144 MHz the central emission is dominated by the S-shaped jets, while the core is not visible (Fig. \ref{fig:jvlalofar}, panel (e)). The brightest, innermost part of the extended east-west lobes is also visible at $\sim$3 -- 6 $\sigma_{\text{rms}}$. The emission is nearly perpendicular to the north-south jets, and it is detected up to a distance of $\sim6''$ ($\sim$29 kpc) on each side of the core. 
The symmetric morphology of the inner north -- south jets strongly suggests that the whole structure mainly lies in the plane of the sky, i.e. the angle between the jet axis and the line of sight is nearly $90^{\circ}$. We note that the two southern hotspots are brighter by a factor $\sim$1.2 -- 1.7 than the northern ones (from 144 MHz to 9 GHz). If this is caused by relativistic effects (e.g. \citealt{Pearson96}), then the southern jet would be the approaching jet and the northern jet the receding one. From the jet/counter-jet ratio, we estimate an upper limit to the inclination angle of $i\leq84^{\circ}$. This supports the above argument of the jets mainly lying in the plane of the sky.
\par We note the absence in the 5.5 GHz JVLA map of the faint E-W jets detected at $\sim4-5\sigma_{\text{rms}}$ in the narrow-band 4.8 GHz VLA images of \citet{Gitti06,Gitti13}. 
The small E-W jet-like features in \citet{Gitti13} may represent the brightest patch of the diffuse, large lobe emission that is visible in the 3 GHz (\texttt{robust = 2}) and in the LOFAR maps. Given the lower sensitivity to extended emission of the \texttt{robust = 0} JVLA maps, such diffuse components are likely resolved out. Alternatively, it is possible that the features in the old VLA maps \citep{Gitti06,Gitti13} were caused by serendipitous noise structures near the core.

  %\begin{SCfigure*}
  \begin{figure*}
   \centering
   \includegraphics[width=\hsize]{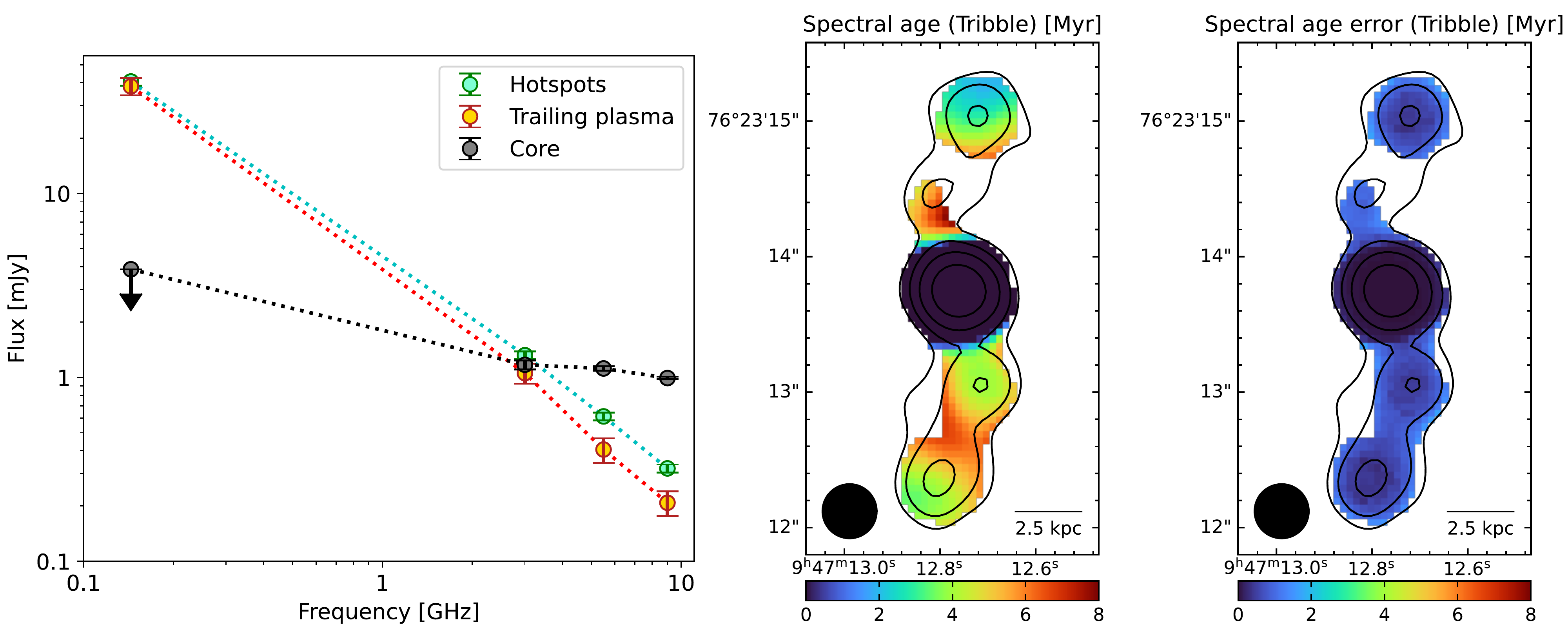}
      \caption[Spectra and radiative ages of the inner jets]{ Spectra and radiative ages of the inner jets. \textit{Left panel:} Radio spectrum of the different components (core, hotspots, and trailing regions) of the north - south radio emission on kpc scales. The total radio spectrum of the hotspots has been obtained by summing the flux densities of the four hotspots within beam-sized regions. The 144 MHz flux density of the core is interpreted as an upper limit, since the core is not detected at low frequency (see \cref{subsubsec:specjets} for details). Note that the dotted lines connecting the points do not represent fits to the spectrum. \textit{Right panel:} Spectral age map (assuming a Tribble model) and associated uncertainty map. Contours from the 3 GHz map used for the spectral index computation (Fig. \ref{fig:jvlalofar-spix}) are displayed on top of the image. The beam is shown in the bottom left corner. See \cref{subsubsec:specjets} for details. 
              }
         \label{fig:jetage}
    \end{figure*}
   %\end{SCfigure*}

\subsubsection{Spectral properties of the inner jets}\label{subsubsec:specjets}
The spectral properties of the radio emission provide useful information on the history of the AGN outbursts. The available JVLA and LOFAR data with similar resolution enable a spatially-resolved spectral study of the north -- south jets over nearly two orders of magnitude in frequency (144 MHz -- 9 GHz). To this end, we produced images at the four frequencies by matching the uv-range of the data (15.2 k$\lambda$ --  481.9 k$\lambda$, corresponding to angular scales of 13.6$''$ -- 0.4$''$) and selecting a uniform weighting scheme. The resulting maps have a resolution of 0.4$''$. 
\par We show in Fig. \ref{fig:jvlalofar-spix} the maps of spectral index between 144 MHz -- 3 GHz (\textit{left}), 3 GHz -- 5.5 GHz (\textit{center}), and 5.5 GHz -- 9 GHz (\textit{right}). Across the four frequencies, the core spectral index remains fairly flat. Using a beam-sized region, and treating the LOFAR flux density as a limit (the core is undetected in the 144 MHz image), we fitted the core measurements with a power-law, finding $\langle\alpha_{core}\rangle = 0.19\pm0.02$.
\par The spectral index along the jets is steeper than in the core, and across the four frequencies ranges between $0.9 \leq \alpha \leq 2$. The spectral index distribution is not uniform, but we note a flatter $\alpha$ at the center of the hotspots, and a progressive steepening behind them. This is consistent with particles being accelerated at the hotspots of jets, which have a spectral index at lower frequencies that is close to the injection index (e.g., \citealt{ODea09}). The flatter hotspot spectrum is evident from Fig. \ref{fig:jetage} (\textit{left panel}). In this figure we compare the core radio spectrum, the total hotspots spectrum (obtained by summing the flux density within beam-sized regions), and the spectrum of the trailing plasma in the remaining area. The difference between the hotspots and the trailing plasma becomes more evident at increasing frequencies. Such spectral steepening at GHz-frequencies is likely caused by radiative losses from particle ageing.  Specifically, the synchrotron energy losses of a population of electrons are faster at higher energy, producing increasingly curved radio spectra over time. The time $t$ that has passed since injection can be recovered from the radio spectrum by measuring the break frequency, $\nu_{b}$ (which describes the increasing curvature), and the magnetic field, $B$, by considering that (e.g., \citealt{slee2001}): 
\begin{equation}
\label{eq:synch}
    t = 1.6\times10^{3} \frac{B^{0.5}}{(B^{2} + B^{2}_{\text{CMB}})}\frac{1}{[\nu_{b}\times(1+z)]^{0.5}} \,\,\text{Myr},
\end{equation}
\noindent where $z$ is the redshift and $B_{\text{CMB}}=3.25(1+z)^{2}$\,[$\mu$G].
\par To investigate the spectral ageing of the relativistic particles, we derived a radiative age map of the inner jets by combining the images at the different frequencies (144 MHz, 3 GHz, 5.5 GHz and 9 GHz). We used the Broadband Radio Astronomy ToolS software (BRATS, \citealt{Harwood13}), that can fit ageing models to 2D maps of radio emission. We tested two radiative age models that consider synchrotron and inverse-Compton losses, where the electron population has a continuously isotropized pitch angle distribution. The JP model \citep{JaffePerola73} uses a constant magnetic field, while the Tribble model \citep{Tribble93} assumes a Gaussian distribution for the magnetic field. We refer the reader to \citet{hardcastle2013,Harwood13} and to the BRATS Cookbook\footnote{\url{https://www.askanastronomer.co.uk/brats/}} for details on the equation of the above ageing models.
\par The input parameters for the JP and Tribble models are the magnetic field B and the injection index $\Gamma$. The magnetic field has been estimated from the 3~GHz map adopting equipartition assumptions (e.g., \citealt{govoni2004}). We measured a k-corrected radio power of $(1.79\pm0.09)\times10^{24}$~W/Hz from a prolate ellipsoid of major semi-axis 1.83'', minor semi-axis 0.66'', and position angle 90$^{\circ}$ from the west axis (volume of 3.34 arcsec$^{3}$, or 392.3 kpc$^{3}$). Using these parameters and assuming an energy ratio of protons to electrons of 1, we derived an equipartition magnetic field of $B_{eq} = 16\pm2\,\,\mu$G along the $\sim$8 kpc -- long jets, and we fixed $B = B_{eq}$. Using the \texttt{findinject} task of BRATS and excluding the region of the core (e.g., \citealt{Brienza20}), we found an injection index $\Gamma = 0.9$. Typical values can range between 0.5 -- 0.8 (e.g., \citealt{Carilli91}), with steeper indices (up to 1.2) found for cluster -- central powerful radio sources (e.g., \citealt{Harwood15}). 
\par Fitting the two models to the data resulted in the Tribble model having a lower reduced-$\chi^{2}=1.12$ than the JP model ($\chi^{2}=1.31$). Moreover, the two models give consistent best-fit radiative ages.
As a note of caution, we report that fixing the magnetic field to the the minimum energy loss field (e.g., \citealt{deGasperin17}) value of $B = 3.4\,\mu$G \citep{Ubertosi21apj} returns radiative ages that are larger than those obtained by adopting $B = B_{eq}$ by $\sim$30\%, but consistent within $\sim2\sigma$.
Considering this, and the fact that the Tribble model describes a more general case, we adopt this one as the best fit to the radio image. We show in Fig. \ref{fig:jetage} (\textit{right panel}) the radiative age map and the associated uncertainty map.
\par Our wide frequency range, which extends over three orders of magnitude, ensures high accuracy on the fitted radiative ages, that have relative uncertainties between 5\% and 25\%. The core has a radiative age consistent with zero, which is in agreement with its spectral index being below the injection index. The jets have radiative ages that range between 2 -- 7 Myr, with the hotspots being younger than the trailing radio plasma behind and around them. In particular, the average age of the hotspots is $t_{\text{rad,H*}} = 4.4\pm0.5$ Myr, compared to the average (hotspot-removed) jet age of $t_{\text{rad,j}} = 5.8\pm0.3$ Myr (see Tab. \ref{tab:lobefit}). 
\par The typical model of an ageing radio galaxy would predict the lower ages to be found in the hotspot of each lobe (or in the proximity of the jet termination point), with the surrounding plasma displaying larger radiative ages (e.g., \citealt{Carilli91}). In the case of RBS~797 we instead see two bright spots for each jet, and a similar radiative age across the four hotspots. This supports the presence of multiple acceleration sites along the jets, not only the terminal ones. We further discuss this finding in \cref{sec:discussion}. 
%%%%%%%%%%%%%%%%%%%%%%%%
\begin{table*}[ht!]
    \centering
    \renewcommand{\arraystretch}{1.5}
    \caption{Summary of spectral age fitting of the different structures of the radio galaxy in RBS~797.}
\begin{tabular}{c|c|c|c|c|c|c|c|c}
\hline
 (1) & (2)  & (3) & (4) & (5) & (6) & (7) & (8) & (9)\\
\hline
          Structure & Component & Model & $B$ & $\nu_{b}$  & $\Gamma$ & $t_{age}$   & $t_{on}$     & $t_{off}$ \\
          & & & [$\mu$G] & [GHz]      &          & [Myr]       & [Myr]        & [Myr]  \\
 \hline
\multirow{2}{*}{\color{black}Inner jets}  & Hotspots  & \multirow{2}{*}{\color{black}Tribble} & \multirow{2}{*}{\color{black}16 (=$B_{eq}$, fixed)} & --&   \multirow{2}{*}{\color{black}0.9 (fixed)}   &$4.4\pm0.5$ & -- & -- \\
& Trailing plasma  &  & & --&     &$5.8\pm0.3$ & -- & -- \\
 \hline
\multirow{2}{*}{\color{black}Outer lobes} & E -- W & \multirow{2}{*}{\color{black}CI-off} & \multirow{2}{*}{\color{black}3.4 (=$B_{min}$, fixed)}& $3.1\pm0.2$& $0.54\pm0.04$     &$31.4\pm6.6$ & $11.4\pm2.4$ & $20.1\pm4.2$ \\
%\hline
& N -- S  & &  & $3.0\pm0.3$& $0.54\pm0.05$     &$32.1\pm9.9$ & $12.8\pm4.0$ & $19.3\pm6.0$ \\
 \hline
\end{tabular}
 \label{tab:lobefit}
 \tablefoot{In the first row, we report the summary of the Tribble model fit to the inner jets of the radio galaxy (see Fig. \ref{fig:jetage} and \cref{subsubsec:specjets}). Note that the fitting has been performed in two dimensions, and the {\it Hotspots} and {\it Trailing plasma} components are average estimates derived from the 2D maps shown in Fig. \ref{fig:jetage}. In the second row, we report the best fit parameters for the CI-off model fitted to the radio spectrum of the E-W and N-S lobes of RBS~797 (see Fig. \ref{fig:lobeage} and \cref{subsubsec:speclobes}). In this second case, the fitting has been performed over the integrated spectra of the radio lobes (bottom panel of Fig. \ref{fig:lobeage}). (1) Structure of the radio galaxy; (2) sub-component of the structure; (3) model used to fit the synchrotron spectra; (4) magnetic field adopted during the fitting procedure; (5) break frequency of the spectrum; (6) injection index; (7) total age of the radio emission; (8) active phase duration (CI-off model only); (9) remnant phase duration (CI-off model only).}
\end{table*}
%%%%%%%%%%%%%%%%%%%%%%%%
%%%%%%%%%%%%%%%%%%%%%%%%
       \begin{figure}[ht!]
   \centering
   \includegraphics[width=\hsize]{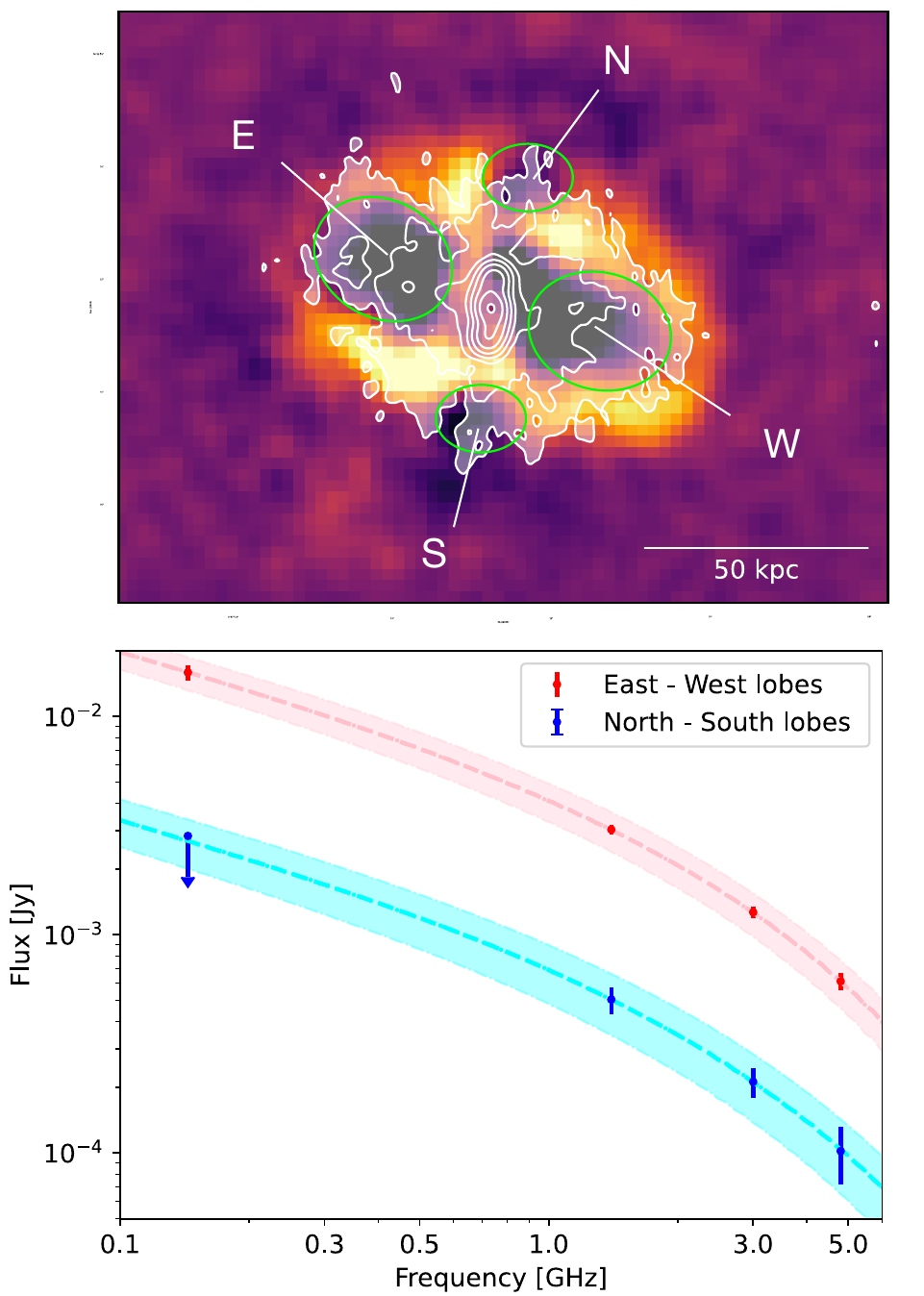}
      \caption[Spectral age study of the perpendicular radio lobes.]{Spectral age study of the perpendicular radio lobes. \emph{Top}: Residual $Chandra$ image of RBS~797, with overlaid white contours from the JVLA 3 GHz image of Fig. \ref{fig:jvlalofar}. The X-ray cavities described in \citet{Ubertosi21apj} are shown with green ellipses and labelled. \emph{Bottom}: Spectrum of the radio emission from the E-W and N-S cavities (see regions in top panel) in RBS~797, fitted with a CI-off model (see \cref{subsubsec:speclobes} for details). Colored areas show the 1$\sigma$ uncertainties on the best-fit model.
              }
         \label{fig:lobeage}
   \end{figure}
   %%%%%%%%%%%%%%%%%%%%%%%%
\subsubsection{Spectral properties of the perpendicular lobes}\label{subsubsec:speclobes}
An open question on RBS~797 is which pair of the perpendicular radio lobes and corresponding X-ray cavities is older. Based on deep X-ray observations, \citet{Ubertosi21apj} determined that the two pairs of X-ray cavities are $\approx$30 Myr old, and they put an upper limit on the age difference between the two outbursts of $\leq$10 Myr. \citet{Ubertosi21apj} also attempted a spectral modeling of the lobes radio spectra, finding a tentative upper limit on the radiative age of $t_{EW}\leq37$~Myr and $t_{NS}\leq38$~Myr. However, given the availability at the time of two frequencies only (1.4 GHz and 4.8 GHz old VLA data), no conclusive result could be obtained. 
\par With the new LOFAR and JVLA data at 144 MHz and 3 GHz, respectively, in combination with the old VLA data published in \citet{Gitti06,Gitti13}, we can now perform a detailed spectral modeling of the radio lobes. Flux densities have been measured from maps obtained by setting a common uv-coverage (4.2 k$\lambda$ -- 175.4 k$\lambda$), using uniform weighting, and a matching resolution of 1.5$''$.
\par We note that the missing uv-coverage in the LOFAR data between baselines of roughly 50k$\lambda$ -- 75k$\lambda$ limits the sensitivity to extended emission between scales of 2$''$ -- 4$''$ \citep{Morabito22}. For this reason, for the fainter N-S lobes we only have upper limits on the extended emission from the LOFAR data. Similarly, the 4.8 GHz VLA data detect the lobes only at $\sim3\sigma_{\text{rms}}$. As the signal-to-noise would be too low, we cannot derive spatially-resolved spectral measurements, but we can obtain integrated values for each component. To this aim, we measured the flux densities within the regions defining the X-ray cavities (see Fig. \ref{fig:lobeage}, top panel). The resulting spectrum of the E-W lobes and the N-S lobes is shown in Fig. \ref{fig:lobeage}, bottom panel.
\par We note that the spectra of the two lobe pairs follow a similar steepening trend towards higher frequencies, with the N-S lobes being a factor of approximately 5 fainter than the E-W lobes. Interestingly, the E-W cavities are also roughly five times more powerful than the N-S cavities, in agreement with the already known scaling between radio and mechanical powers (e.g. \citealt{Birzan08,Cavagnolo10,OSullivan11}).
\begin{comment}
\begin{table}
    \centering
    \renewcommand{\arraystretch}{1.5}
    \caption{Best fit parameters for the CI-off model fitted to the radio spectrum of the E-W and N-S lobes of RBS~797 (see Fig. \ref{fig:lobeage} and \cref{subsubsec:speclobes}). (1) Pair of radio lobes; (2) break frequency; (3) injection index; (4) total age of the radio emission; (5) active phase duration; (6) remnant phase duration.}
\begin{tabular}{c|c|c|c|c|c}
\hline
 \textbf{(1)} & \textbf{(2)}  & \textbf{(3)} & \textbf{(4)} & \textbf{(5)} & \textbf{(6)} \\
\hline
          & $\nu_{b}$  & $\Gamma$ & $t_{age}$   & $t_{on}$     & $t_{off}$ \\
          & [GHz]      &          & [Myr]       & [Myr]        & [Myr]  \\
 \hline
E-W  & $3.1\pm0.2$& $0.54\pm0.04$     &$31.4\pm6.6$ & $11.4\pm2.4$ & $20.1\pm4.2$ \\
\hline
N-S  & $3.0\pm0.3$& $0.54\pm0.05$     &$32.1\pm9.9$ & $12.8\pm4.0$ & $19.3\pm6.0$ \\
 \hline

\end{tabular}
 \label{tab:lobefit2}
\end{table}
\end{comment}
\par Using the software \texttt{SYNCHROFIT} \citep{Quici22}, which is suited to analyze integrated spectra, we fitted the two spectra using a continuous injection with a remnant phase model (CI-off, \citealt{Komissarov94}). With respect to the JP or Tribble models (that are based on a single injection event), the CI-off model describes an electron population that is injected at a constant rate for a duration $t_{on}$ (the active phase), then the nuclear activity shuts down and the electrons age for a time $t_{off}$ (the remnant phase). 
During the active phase, the radio spectrum has a curved shape, described by a break frequency $\nu_{b}$, which depends on the total age of the radio source as in Eq. \ref{eq:synch}. For $\nu\geq\nu_{b}$, the spectral index is $\Gamma+0.5$, where $\Gamma$ is the injection spectral index. Then, at a time $t_{on}$, the source injecting fresh electrons switches off and the synchrotron-emitting electrons continue ageing for a time $t_{off}$ (the remnant phase). This produces a second break frequency in the spectrum, $\nu_{b,2}$, defined as:
\begin{equation}
    \nu_{b,2} = \nu_{b}\times\left(\frac{t_{age}}{t_{off}}\right) = \nu_{b} \times\left(\frac{t_{off}+t_{on}}{t_{off}}\right)\,.
\end{equation}
For details on this model and the corresponding equations, see \citet{Komissarov94}; see also \citet{Biava2021} for a similar application. 
\\The choice of the CI-off for the lobes of RBS~797 is physically motivated by the fact that the diffuse lobe emission is not currently powered by active jets, as the most recent activity is in the north - south direction and on smaller scales (see \cref{subsec:kpc}). Indeed, we found that single injection models are unable to correctly fit the data.
\par The model requires the redshift (z = 0.354) and the magnetic field to be fixed, and returns the injection index $\Gamma$, the break frequency $\nu_{b}$ (which is related to the total age of the source, $t_{off}+t_{on}$), and the remnant fraction ($t_{off}/(t_{off}+t_{on})$). In the region of the lobes, we measure an equipartition magnetic field that ranges between $3.2\,\mu$G to $4.2\,\mu$G depending on the selected region. For comparison, the minimum energy loss field is $B_{min} = 3.4\,\mu$G. Given the similarity between these values, we adopted a common magnetic field $B = B_{min}$ during fitting. 
\par As it is possible to see from Fig. \ref{fig:lobeage}, the CI-off model provides a good description of the integrated radio spectrum of the E-W and N-S lobes ($\chi^{2}$/d.o.f. of 1.03 and 1.09, respectively). The best fit parameters for the E-W and N-S lobes are reported in Tab. \ref{tab:lobefit}. We find that the spectra of the two pairs of lobes are described by similar parameters. First, we note that the total age of the radio lobes ($\sim$30 Myr) is in good agreement with the age of the X-ray cavities \citep{Ubertosi21apj} and of the corresponding weak shock in the ICM \citep{Ubertosi23apj}. Second, the similar total age, active phase duration, and remnant phase duration of the E-W and N-S lobes suggest that it is difficult to determine which outburst is younger. Rather, these results support a coeval origin of the two pairs of radio lobes, as suggested before from the X-ray data. We discuss these results in the framework of radio activity in RBS~797 in \cref{sec:discussion}. 
  \begin{figure*}[ht!]
  \sidecaption
   \centering
   \includegraphics[width=0.7\linewidth]{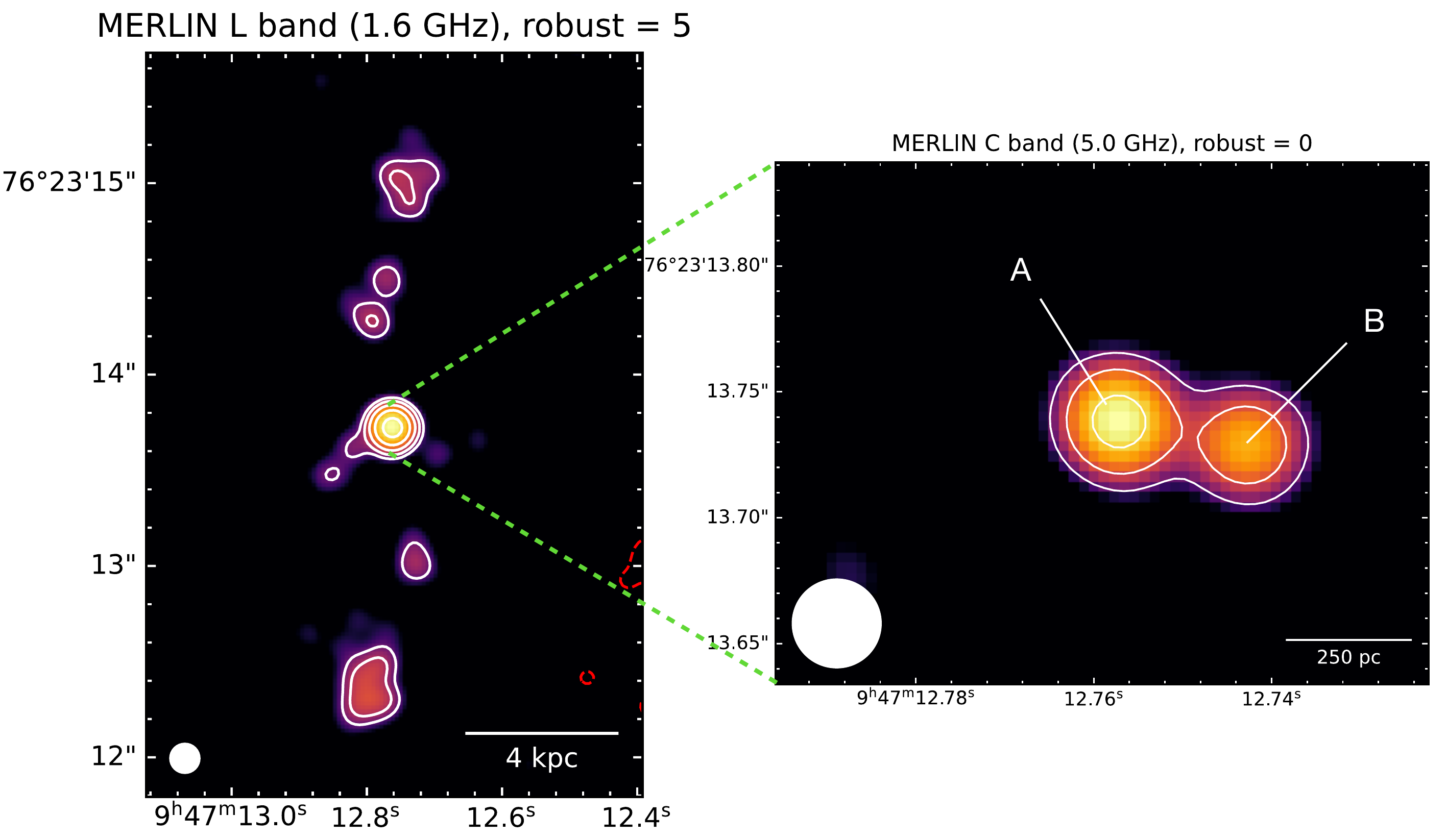}
      \caption[e-Merlin images at 1.6 GHz (left) and 5 GHz (right) of RBS~797.]{e-Merlin images at 1.6 GHz (left) and 5 GHz (right) of RBS~797. In both panels, contours are drawn at $5\sigma_{\text{rms}}$, and increase by a factor of 2; the first negative contour (at $-5\sigma_{\text{rms}}$) is drawn in dashed red (not visible in the right panel). The beam is shown in the bottom left corner (0.15$''\times$0.15$''$ at 1.6~GHz and 0.035$''\times$0.035$''$ at 5~GHz). See Tab. \ref{tab:imagepar} for details on the resolution and noise level of the images. Structures discussed in this work are labelled.}
         \label{fig:merlin}
   \end{figure*}
   
\subsection{From the kpc to the pc scale}\label{subsec:kpcpc}
We present in Fig. \ref{fig:merlin} the images obtained from the e-Merlin observations of RBS~797 at L band (1.6 GHz) and C band (5 GHz). Tab \ref{tab:imagepar} reports the parameters used during imaging and the corresponding image properties. 
\par The 1.6 GHz image shows the inner jet structure with a resolution of 0.15$''$ ($\sim$0.7 kpc). In particular, the core is visible at the center of the image (with an unresolved flux density of 1.51 mJy), and the jet emission mainly coming from the four hotspots is recovered on larger scales (total flux density 3.4 mJy). This image represents the highest resolution view of the $\sim$8 kpc long jets of RBS~797, and confirms the S-shaped morphology revealed by the JVLA and LOFAR images of Fig. \ref{fig:jvlalofar}. 
\par The 5 GHz image has a resolution of 0.035$''$ ($\sim$170 pc). The emission is resolved into two components, with the brighter one (labelled A) centered on the position of the radio core (RA, DEC = 09:24:12.76, +76:23:13.74) and the fainter one (labelled B) located at a distance of 0.05$''$ ($\sim$250 pc) westward. Although the dynamic range of the 5 GHz e-Merlin image is relatively small (around 25), the two components are detected at more than 5$\sigma_{\text{rms}}$, up to $\sim$28$\sigma_{\text{rms}}$ for the brighter component and $\sim$20$\sigma_{\text{rms}}$ for the fainter one. 
Using the software \texttt{DIFMAP} we fitted the combination of two Gaussian components to the uv data. We found that A is described by a Gaussian with flux density $S_{A} = 0.55\pm0.04$ mJy and a nominal deconvolved major axis of $1.1\pm0.1$ mas ($5.4\pm0.5$ pc). This is a factor of $\sim$32 smaller than the beam size, indicating that the component is unresolved. The second component is described by a Gaussian with flux density $S_{B} = 0.38\pm0.04$ mJy and a deconvolved major axis of $10.2\pm1.4$ mas ($51\pm7$ pc). This is a factor $\sim$3 smaller than the beam size, possibly suggesting that the component may be extended.
\par Given its position, compactness, and flux density, we identify the brighter component, A, with the central engine of RBS~797. The second component is more difficult to interpret. This fainter component, B, could arise from a second SMBH in the BCG of RBS~797, which has been argued as a possible explanation for the multiple and misaligned outbursts in this cluster \citet{Gitti06,Gitti13,Ubertosi21apj}. Alternatively, it could represent remnant extended emission from the large scale outburst (see \cref{subsec:kpc}), since its relative position with respect to the core matches the orientation of the large east -- west lobes of the radio galaxy (see Fig. \ref{fig:jvlalofar}). This alternative interpretation would be in agreement with its possible classification as an extended component from the \texttt{DIFMAP} fitting procedure. We also considered the possibility that this feature would be part of the southern approaching jet, however this seems hard to reconcile with the position angle of the
component S1 detected with VLBA (see next section). We exploit the information from the VLBA and EVN data presented in the next section to discuss which scenario is more plausible. 
\begin{figure*}[ht]
   \centering
   \includegraphics[width=\hsize]{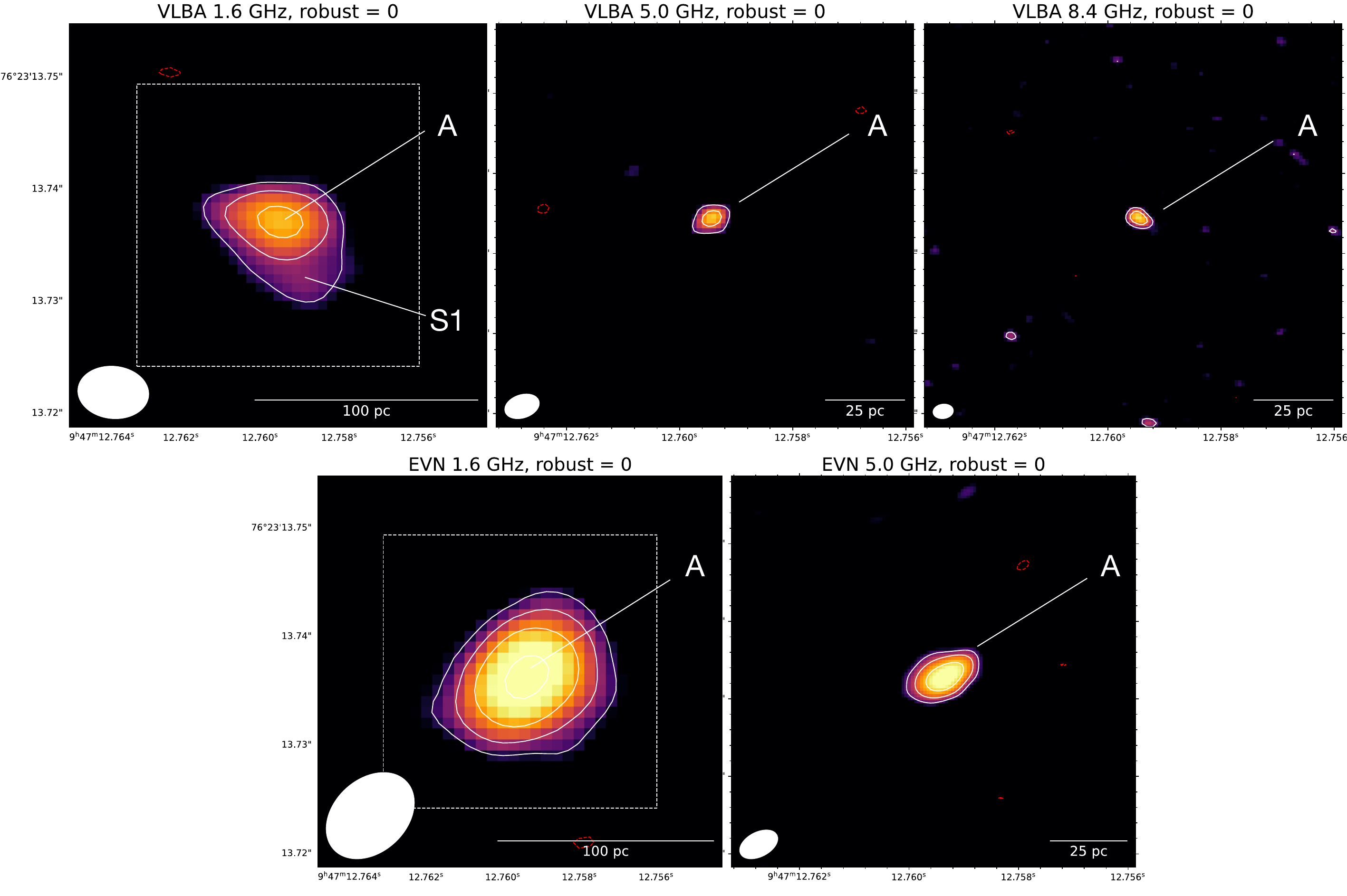}
      \caption[VLBA and EVN maps of RBS~797 at different frequencies.]{VLBA and EVN maps of RBS~797 at different frequencies. {\it Top panels}: VLBA images at 1.6 GHz, 5 GHz, and 8.4 GHz. The beam sizes are 6.3$\times$4.6 mas at 1.6~GHz, 2.2$\times$1.4 mas at 5~GHz, and 1.3$\times$0.9 mas at 8.4~GHz. {\it Bottom panels}: EVN images at 1.6 GHz and 5 GHz. The beam sizes are 9.1$\times$6.7 mas at 1.6~GHz, and 2.6$\times$1.6 mas at 5~GHz. In each panel, the beam is shown in the bottom left corner, and contours start at $5\sigma_{\text{rms}}$ and increase by a factor of 2; the first negative contour at $-3\sigma_{\text{rms}}$ is drawn in dashed red (see Tab. \ref{tab:imagepar} for the resolution and noise levels of the images). The 1.6 GHz VLBA and EVN images have a matched field of view. The white dashed square shows the extent of the images at 5 GHz (EVN and VLBA) and at 8.4 GHz (VLBA). Structures discussed in this work are labelled.
              }
         \label{fig:vlbaevn}
   \end{figure*}
   \begin{table*}[ht]
    \centering
    \caption{Results of \texttt{modelfit} in \texttt{DIFMAP} applied to EVN and VLBA data of RBS~797.}
    \renewcommand{\arraystretch}{1.5}
\begin{tabular}{c|c|c|c|c|c|c|c|c|c|c}
\hline
 (1) & (2)  & (3) & (4) & (5) & (6) & (7) & (8) & (9) & (10) & (11)\\
\hline
Telescope & Project Code  & Frequency & Obs. Date & Robust & $\chi^{2}/D.o.f.$ & Component & a & b/a & $\Phi$ & S\\
 & & [GHz] & & & & & [mas] & & [$^{\circ}$]& [mJy] \\ 
 \hline
 \multirow{5}{*}{\color{black}VLBA}   &  \multirow{4}{*}{BG~224}    &     \multirow{2}{*}{1.6} & \multirow{2}{*}{Feb. 4, 2014} &\multirow{2}{*}{\texttt{0}} & \multirow{2}{*}{1.13 (229130)} & A & 2.5 & 1.0 & 0 &  0.75$\pm$0.09 \\
   &     &       &  &  & & S1 & 5.8 & 0.7 & -16.1 &  0.27$\pm$0.04 \\
 \cline{3-10}
      &   &      5.0  & Feb. 4, 2014 & \texttt{0} & 1.14 (54424) & A & 0.6 & 1.0 & 0 & 0.47$\pm$0.05  \\
   \cline{3-10} 
   &   &      8.4 & Feb. 4, 2014 & \texttt{0} & 1.19 (109548) & A & 0.05 & 1.0 & 0 & 0.40$\pm$0.05 \\

   \cline{2-10} 
      & BE~056    &      5.0 & Jul. 4, 2014 & \texttt{3} & 1.21 (13831) & A & 2.5 & 1.0 & 0 & 0.47$\pm$0.06  \\

\hline
\multirow{4}{*}{\color{black}EVN}     & \multirow{2}{*}{EG080}   &      1.6 & Feb. 26, 2014 & \texttt{0} &1.16 (236740)& A & 3.0 & 1.0 & 0 & 1.65$\pm$0.17 \\

\cline{3-10}
     &    &      5.0 & Mar. 9, 2014 & \texttt{0} & 1.07 (314572) & A & 0.9 & 1.0 & 0 & 0.70$\pm$0.07 \\

\cline{2-10}
     & RSG05     &      5.0 & May 3, 2013 &  \texttt{0}& 1.12 (9068) & A & 3.4 & 1.0 & 0 & 0.84$\pm$0.09 \\
\cline{2-10}
    & RSG07     &      1.6 & Dec. 1, 2015 & \texttt{0} & 1.06 (12092)& A & 7.0 & 1.0 & 0 & 1.91$\pm$0.22\\

\hline

\end{tabular}
\label{tab:difmap}
\tablefoot{(1) Name of the radio telescope; (2) project code; (3) observing frequency; (4) date of the observations; (5) \texttt{robust} parameter used during imaging; (6) $\chi^{2}$/degrees of freedom; (7) label of the component; (8) major axis of the component, in mas; (9) ratio between minor and major axis of the component; (10) position angle of the ellipse describing the component, in degrees; (11) flux density of the component. See \cref{subsubsec:vlbaevn} for details. }
\end{table*}

%UVRANGE MERLIN : 0.2 - 4 MEGALAMBDA
%UVRANGE EVN : 1.2 - 7.6 megalambda

\subsection{The pc scale}\label{subsec:pc}
We show in Fig. \ref{fig:vlbaevn} the VLBA images at 1.6 GHz, 5 GHz and 8.4 GHz (top panel), and the EVN images at 1.6 GHz and 5 GHz (bottom panel) at mas resolution. The parameters of the images are reported in Tab. \ref{tab:imagepar}. From these images we identify a single source at the phase center; the emission appears unresolved, with the exception of the VLBA 1.6 GHz image, where a faint extension is visible south of the peak. 
\par We used \texttt{DIFMAP} to identify components from the uv-data and measure their flux densities. Based on this analysis, we found that the VLBA data at 1.6 GHz are well described by the combination of two components: a bright one (A, 0.75$\pm$0.09 mJy) coincident with the radio peak, and a fainter one (S1, 0.27$\pm$0.04 mJy) at a distance of 12 mas ($\sim$60 pc) south of A. We observe that the position angle of the jet-like component S1 matches the jet orientation on JVLA and LOFAR scales (see Fig. \ref{fig:jvlalofar}). Additionally, the one-sidedness of the jet is consistent with the possible interpretation presented in \cref{subsubsec:specjets} of the southern jet being the approaching one. The VLBA data at higher frequencies are described by a single, unresolved component (coincident with A), with a flux density of 0.47$\pm$0.05 mJy (at 5 GHz) and 0.40$\pm$0.05 mJy (at 8.4 GHz). 
\par The EVN data at 1.6 GHz is fitted with a single component with a flux density of 1.65$\pm$0.17 mJy, that we identify as A from the VLBA maps. Similarly, the 5 GHz data are fitted with a single component with flux density 0.70$\pm$0.07 mJy. Component S1 is not visible in the 1.6 GHz EVN map, although there are hints of a south-west extension in the map shown in Fig. \ref{fig:vlbaevn} (bottom left). Additionally, the total flux density within the $5\sigma_{\text{rms}}$ contour at 1.6 GHz is higher than the peak flux by $\sim$20\% (see Tab. \ref{tab:imagepar}), further supporting the presence of extended flux in the EVN map. Therefore, A$+$S1 are likely unresolved into a single component (the EVN beam area is twice as large as the VLBA beam area). However, forcing a smaller beam to the EVN data and using a uniform weighting to improve the resolution results in a worse image sensitivity, which prevents us from fitting other components to the data in \texttt{DIFMAP}.
\par The simultaneous VLBA observations at the different frequencies (all performed on February 4, 2014) allow us to obtain a reliable measurement of the source spectral index. We measured the flux density of the core component (A) from maps obtained by matching the uv-range between the observing frequencies (5~M$\lambda$ -- 50~M$\lambda$), and we fitted a power-law to the flux densities at 1.6 GHz, 5 GHz, and 8.4 GHz. This returned a spectral index of $\alpha_{A} = 0.25\pm0.04$. Such rather flat value is consistent with the core spectral index on the kpc scale from LOFAR and JVLA data of $0.19\pm0.02$ (see \cref{subsec:kpc}). We also deduce that the spectral index of component S1 is steeper than 0.65 at 3$\sigma$ confidence, based on the sensitivity of the 5 GHz and 8.4 GHz observations (which do not detect this component). 
\linebreak
\\ Comparing our results with those presented in \citet{Gitti13}, we find the absence in our maps of the second component tentatively detected in the snapshot EVN observation at 5 GHz (performed on May 3, 2013) at a distance of 77 pc south-west from the radio core. The non-detection at all frequencies and between the different instruments suggests that it was, most likely, a spurious detection. A re-analysis of the data indicates that the component was in fact an artifact caused by the phase calibrator being resolved into a double itself.
\par Ultimately, we note that from the VLBA and EVN data we find no evidence of the second component B visible in the e-Merlin image at 5 GHz of Fig. \ref{fig:merlin}. This supports our interpretation of B as extended radio emission, likely resolved out in the VLBI data due to the different uv-coverage. 

\subsection{Does the radio core of RBS~797 show time variability?}\label{subsubsec:variab}
We note that there is an inconsistency in the flux density of RBS~797 between the different observations. At 1.6 GHz, the EVN total flux density is larger than the VLBA total flux density by a factor of $\sim$1.75, while at 5 GHz the EVN total flux density is larger than the VLBA total flux density by a factor of $\sim$1.5. This is inconsistent with the typical flux density scale error associated with the EVN and VLBA telescopes (around 10\%).
\par To gain more insights on this difference, we retrieved other EVN and VLBA snapshot observations of RBS~797 (see \cref{subsec:archive} and Tab. \ref{tab:sumobs}) and we measured the (unresolved) flux densities in \texttt{DIFMAP}. 
Based on the flux densities reported in Tab. \ref{tab:difmap}, we find that the 5 GHz emission oscillated between $\sim$0.7 mJy and $\sim$0.5 mJy between May 2013 and July 2014. The 1.6 GHz flux density increased from $\sim$1 mJy to $\sim$1.7 mJy between the 4$^{th}$ and the 26$^{th}$ of February 2014, and the following observation on December 1, 2015 measured a flux density of $\sim$1.9 mJy (consistent with the previous measurement). 
A clear trend is that the EVN flux densities are always higher than the corresponding VLBA flux densities at the same frequency. The 5 GHz~EVN flux density (0.7~mJy) is also higher than the nearly contemporaneous 5~GHz eMerlin flux density (0.55~mJy). This may indicate a technical origin, or a different sensitivity to the structures of the source. However, we verified that matching the uv-coverage and image parameters does not solve such inconsistency. Excluding the Jodrell Bank and Westerbork stations from the EVN data (which provide the shortest spacings and therefore the highest sensitivity) still does not change these results. Additionally, the mismatch is not caused by transferring solution from a strongly variable phase calibrator (J0954+7435), which does not show significant flux density variations. 
\\\indent Interstellar scintillation may cause flux density variations in compact radio sources such as flat spectrum AGN, which typically occur on timescales of a few hours or days, at most (e.g., \citealt{Koay18}). Yet, the variations measured in RBS~797 are on weeks and month timescales, and we do not find evidence of significant variability within the duration of each observation. 
\\\indent One possibility is that the variability is intrinsic to the source itself, which is not uncommon for the radio cores of BCGs (e.g., \citealt{Hogan15a,Hogan15b}). For active, long lived radio galaxies, the variation may be caused either by (a) a change in the accretion rate and thus in jet power and brightness; or (b) the launching of distinct knots in the jets; or (c) if the jets are precessing, by the varying Doppler beaming of the ejected jet components. Considering the kpc-scale properties of the radio galaxy in RBS~797 (see \cref{subsec:kpc} and \cref{subsubsec:modelprec}), the above scenarios seem plausible. Alternatively, in the case of young radio sources with recently renewed jets, variability may be attributed to the synchrotron peak progressively shifting toward lower frequencies or to the jets interacting with the surrounding material (e.g., \citealt{bruni2019,wolowska2021}).  If the radio core of RBS~797 is experiencing a fourth cycle of recently restarted AGN activity, then the above mechanism may provide a framework to interpret the possible variability.
\\Interestingly, we find evidence of time variability also on VLA scales. We compared the 4.8 GHz VLA observations published in \citet{Gitti06} (performed in 2004) and our JVLA data at 5.5 GHz (4.5 -- 6.5 GHz bandwidth), obtained in 2022. We built comparable maps by matching the uv-range, frequency, and bandwidth (4.8 GHz, 128 MHz) of the data. Using circles with radius of 1$''$ centered on the radio core, we measured flux densities of $S_{4.8}^{2004} = 1.07\pm0.06$ mJy and $S_{4.8}^{2022} = 1.54\pm0.14$ mJy. The two measurements are different by a factor 1.44 at a significance of $2.9\,\sigma$. Thus, we have independent indications of variability from the radio core of RBS~797, besides the evidence from VLBI data. Nevertheless, given the suspicious systematic difference between the EVN and VLBA observations, and the poor sampling of the light curve, we cannot draw strong conclusions on the origin and amplitude of such variability, which currently remains unexplained. Future ad-hoc monitoring of this source may shed lights on this issue.

\section{Discussion}\label{sec:discussion}
\subsection{Deciphering the morphology of the radio galaxy in RBS~797}
The new radio images shown in the previous sections confirm the earlier claims that the central AGN of RBS~797 has been going through several episodes of activity, the directionality of which has changed over time. Combining our results with the previous works of \citet{Gitti06,Gitti13,Ubertosi21apj,Ubertosi23apj}, we can summarize what is currently known on the outburst history of RBS~797 as follows. On 50 -- 100 kpc scale the presence of three shock fronts and diffuse radio emission (see also \citealt{bonafede2023}) supports an interaction between the AGN and its environment that dates back to $\approx$ 80 Myr ago. On scales of 10 -- 50 kpc there are the perpendicular pairs of radio lobes and corresponding X-ray cavities that have similar ages ($\sim$30 Myr) down to a few Myr of accuracy (\cref{subsubsec:speclobes} and Tab. \ref{tab:lobefit}). Zooming into the inner 10 kpc, north-south curved jets are visible, with multiple sites of relativistic particle acceleration. The S-shaped jets are radiatively young, with an age of a few Myr (\cref{subsubsec:specjets}, \cref{subsec:kpcpc}, and Tab. \ref{tab:lobefit}). Ultimately, going down to hundreds and tens of pc an active core connected to a southward jet feature is visible, with possible evidence of time variability on months timescales (\cref{subsec:pc} \& \cref{subsubsec:variab}).
\par To understand the history of the multiple misaligned outbursts we need to identify the mechanism that drives such changes in orientation. A clear dynamical feature is the S-shaped jet morphology. These shapes have long been associated with precession over time of the jet axis (e.g., \citealt{Rubinur17,Horton20,Nolting23} and references therein), which is likely to be occurring in RBS~797. Indeed, we have independent evidence that the north -- south jets are precessing. First, we noticed the presence of multiple hotspots in the radio maps, which is one of the expected signatures of precession (e.g., \citealt{Mahatma23}). Second,  the hotspots have similar spectral indices and radiative ages, indicating that particle are simultaneously being accelerated at each of these locations (see also \citealt{Hardcastle07}), which can be a consequence of the swirling jet path \citep{Horton23}. Considering the above information, in the next paragraphs we use the observed jet morphology in RBS~797 to constrain the main parameters of the precession motion. 
     \begin{figure}[ht!]
   \centering
   \includegraphics[width=\hsize]{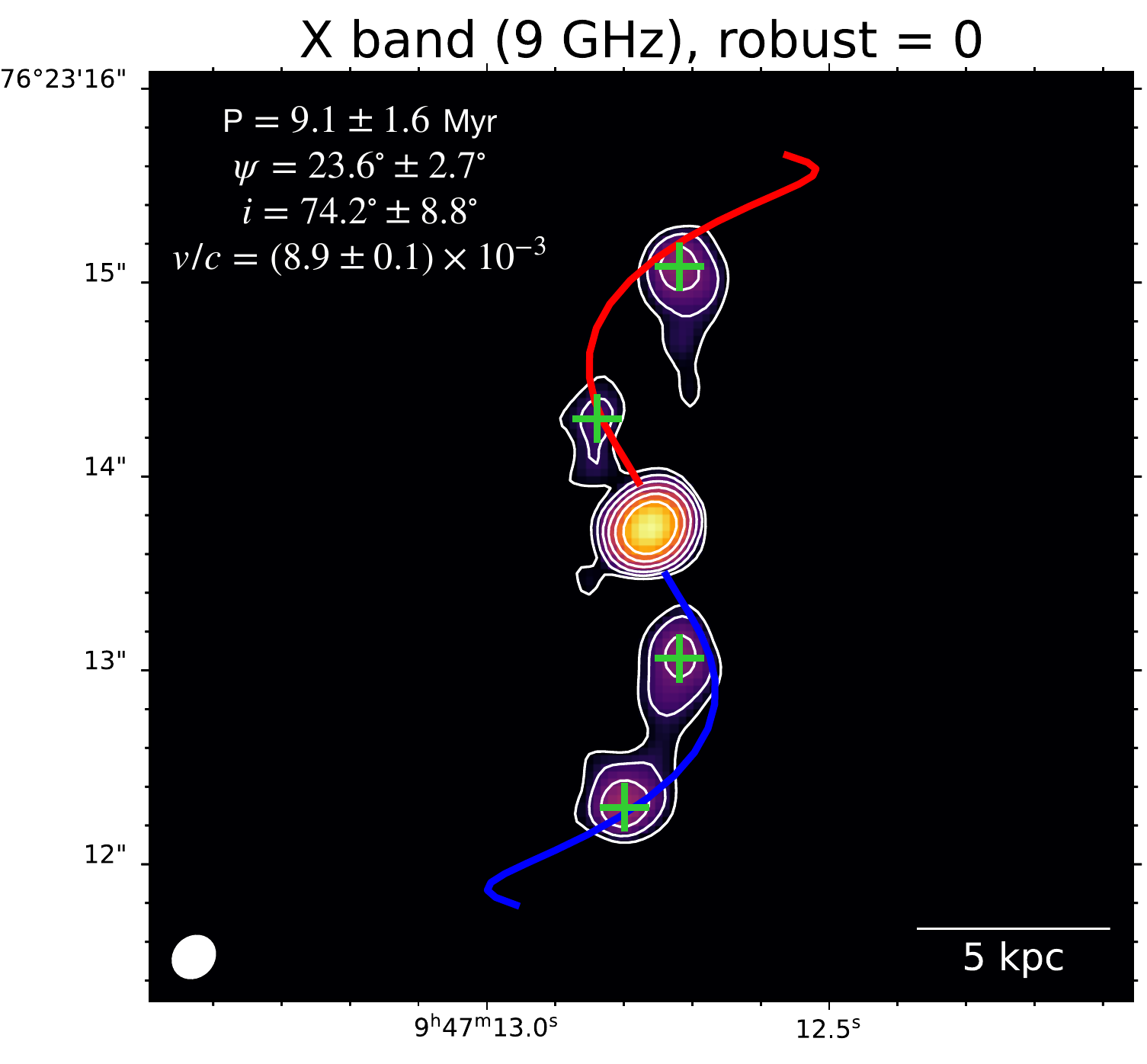}
      \caption[Jet precession in RBS~797]{Jet precession in RBS~797. Precession model of \citet{Hjellming81} fitted to the radio images of the kpc-scale jets in RBS~797. Contours are the same as those shown in Fig. \ref{fig:jvlalofar}. The red line represents the receding jet path and the blue line represents the approaching jet path. The location of the green crosses corresponds to the position of the observed hotspots, while their extent represents the resolution of the radio observation (the beam is shown in the bottom left corner) Best fit parameters are reported in the top left corner. We note that the uncertainties on the best fit parameters represent the combination of statistical and systematic uncertainties (see \cref{subsubsec:modelprec} for details).
              }
         \label{fig:precessmodel}
   \end{figure}
\subsubsection{Modelling jet precession}\label{subsubsec:modelprec}
The distinct S-shape of the kpc-scale jets in RBS~797 visible in Fig. \ref{fig:jvlalofar} may be explained with a precession motion around the jet axis. Such morphologies have been observed in several radio galaxies, from pc to hundreds of kpc scales (e.g., \citealt{liuzzo2009,Machalski16,Bruni21,Nandi21}). The morphological signatures of jet precession can be linked to the dynamical parameters that regulate such motion. These are, mainly, the precession period $P$, the half-opening angle of the jet precession cone $\psi$, the inclination of the jet along the line of sight $i$, and the jet advance speed $v$. The exact mapping between the radio morphology and these quantities depends on the model assumed for the physics of jet propagation and precession. 
\par Here we consider the kinematic precession model of \citet{Hjellming81}, which includes relativistic effects\footnote{While it was built to fit the jet morphology of the microquasar SS433, it does not assume any a-priori condition related to microquasars, and it is relatively simple. Thus, it can be generalized to describe jets from radio galaxies. Indeed, it has been used before to model extragalactic AGN jets (see e.g., \citealt{Rubinur17,Kharb19})}. The model parameters are the distance to the source ($d$), the position angle of the jet precession axis in the plane of the sky ($\chi$), the precession period ($P$), the half-opening angle of the jet precession cone ($\psi$), the precession axis inclination to the line of sight ($i$), and the jet advance speed ($v/c$). From the equations detailed in \citet{Hjellming81}, the model predicts the position at a time $t_{eject} + \Delta t$ of two opposite jet blobs launched from the SMBH at a time $t_{eject}$. The position is expressed in the form of offsets from the core position $\delta RA$ and $\delta DEC$. Evolving the model from the initial time $t_{eject}$ (for example, 20 Myr ago) to $t_{obs}$ (i.e. the present time) produces the predicted model jet path projected onto the plane of the sky over 20 Myr of jet activity. Such model can be readily compared to the real jet path observed from radio maps. 
\par In order to derive the parameters of the jet precession motion in RBS~797, we followed a similar approach to that of \citet{Coriat19}. We used the \texttt{scipy.optimize} package to minimize the difference between (a) the observed offsets of each hotspot (H1, H2, H3 and H4) with respect to the core, $\delta RA_{obs}$ and $\delta DEC_{obs}$, and (b) the jet path predicted by the model in the form $\delta RA_{mod}$ and $\delta DEC_{mod}$. The observed hotspot offsets have been weighted by the area of the radio beam during the minimization process.
\par The model parameters that we fixed before evolving the model are the distance to the source $d = 1881.7$ Mpc (the luminosity distance), and the position angle of the jet precession axis in the plane of the sky, $\chi = 90^{\circ}$ (that is the position angle of the whole S-shaped jet structure, measured from the west axis). We set constraints in input to the model parameters as follows:
\begin{itemize}
    \item $P$: the precession period was bound to vary between 2 Myr (the lowest radiative age we measure from the spectral modelling of the jets, see Fig. \ref{fig:jetage}, \textit{right panel}) and 20 Myr (twice the maximum radiative age we measured). We verified that for $P>20$ Myr the model predicts a jet bending on scales of tens of kpc, which would be in disagreement with the observed bending on a few kpc scale.
    \item $\psi$: the half-opening angle of the jet precession cone was bound to vary between 10$^{\circ}$ and 30$^{\circ}$ to approximately match the observed morphology of the S-shaped jets.
    \item $i$: the jet precession axis inclination with respect to the line of sight was bound between 60$^{\circ}$ and 90$^{\circ}$. As noted in \cref{subsec:kpc}, the symmetry of the jets suggests that the whole structure is not strongly oriented towards the observer. 
    \item $v/c$: the jet advance speed in units of the speed of light $c$ was bound between $10^{-3}c$ and $10^{-1}c$, in comparison with typical speed of kpc -- scale jets (e.g., \citealt{Meyer17,Kappes19,Perucho19}).
\end{itemize}
The minimization procedure consists in \texttt{scipy.optimize} exploring the parameter space of the above quantities and computing every time the predicted jet path. The optimized model is the one for which the distance between the predicted and observed jet paths is the smallest. The minimization was realized 1000 times, and we assumed the dispersion of the derived parameters around the mean as a good estimate of the statistical error $\sigma_{stat}$. Additionally, the minimization was performed not only at one frequency, but using all the available radio maps that detect the hotspots: LOFAR at 144 MHz, JVLA at 3 GHz, 5.5 GHz, 9 GHz (Fig. \ref{fig:jvlalofar}), and e-Merlin at 1.4 GHz (Fig. \ref{fig:merlin}). The dispersion around the mean of the precession parameters at different frequencies was assumed as a good indicator of the systematic error $\sigma_{syst}$. The final uncertainty on each best fit parameter is given by $\sqrt{\sigma_{stat}^{2} + \sigma_{syst}^{2}}$.
\par We show in Fig. \ref{fig:precessmodel} an overlay of the best-fit precession model on the 9 GHz JVLA image of RBS~797, where we also report the best-fit parameters. The predicted jet path is color coded according to the approaching (blue) and the receding (red) jet. We obtain a precession model with period $P \sim 9$ Myr, rotating with a half-opening precession cone angle of $\psi\sim24^{\circ}$ and inclined by $i \sim 74^{\circ}$ degrees with respect to the line of sight. The jet advances with a speed of $v\sim10^{-2}\,c$. We observe that within the uncertainties on the hotspots position (given by the resolution of radio observations), the model provides a nice representation of the jets morphology. The predicted inclination along the line of sight is in line with the jets mainly lying in the plane of the sky. The best-fit jet advance speed agrees with the order-of-magnitude dynamical speed of the jets: based on the radiative age of the jet structures ($\sim$2 -- 7 Myr) and the jet length of $\sim$8 kpc, we would expect a dynamical velocity ranging between $0.004\,c\lessapprox v_{dyn}\lessapprox0.01\,c$. We additionally note that the comparison between the estimated jet radiative ages (\cref{subsubsec:specjets}) and the predicted precession period indicates that the jets are currently half way through a full precession round. 

\subsubsection{The connection between the inner jets and the outer lobes: single or multiple precession modes?}\label{subsubsec:jetlobeconnection1}
The precession model detailed above provides a good description of the north -- south jet motion. However, it is not straightforward to fit the older outbursts into this scenario. The perpendicular pairs of radio lobes on $\sim$50 kpc scales have been inflated over similar timescales, and do not lie on the jet path predicted by the precession model (see Fig. \ref{fig:precessmodel}). Here we discuss whether the morphological evidence support a single dynamical effect, or if the combination of multiple mechanisms is required to explain all the outbursts.
\linebreak
\par Can a single precession mode explain the multi-faceted morphology visible in Fig. \ref{fig:jvlalofar}? Numerical simulations found that a prolonged wobbling outflow coupled with jet self-interactions can cause very complex morphologies to appear (e.g., \citealt{Horton20,Horton23,Lalakos22,Nolting23}). Synthetic radio maps presented in these works show a plethora of structures in multiple directions (even orthogonal ones). \citet{Nolting23} stressed that the viewing angle of the observer ($i$ in this work, $\theta$ in their work) may generate very different apparent radio shapes. 
The case of RBS~797 may be akin to those of the above simulations. However, the most \say{misaligned morphologies} (as those observed in RBS~797) occur either when the half-opening angle of the precession cone is large (up to $\sim$45$^{\circ}$) or when the precession axis is nearly aligned with the line of sight ($i\leq 45^{\circ}$). On the contrary, we found in \cref{subsubsec:modelprec} that the precession cone is relatively narrow ($\psi\sim24^{\circ}$) and that the source is not strongly oriented towards us ($i\sim74^{\circ}$). Thus, the above scenarios may apply to RBS~797 only if the parameters that regulate the precession motion are time dependent (e.g., \citealt{Marti97,Laing14,Giri22}). For example, if the half-opening angle of the precession cone $\psi$ was larger in the past, and possibly as high as 45$^{\circ}$, then it would be possible to find radio emission in the direction perpendicular to that of the current jet axis. Thus, we argue that a simple and stationary precession model cannot explain the whole history of AGN activity. 
\linebreak
\par We can only be speculative about mechanisms that are more elaborate than a single precession mode. A first possibility is that precession was already in place $\sim$30 Myr ago, during the large lobe formation. A short precession period (a few Myr) and a large precession cone half-opening angle ($\psi\approx45^{\circ}$) would explain how the N -- S and E -- W lobes (and the associated cavities) have very similar radiative and dynamical ages (see also \citealt{Ubertosi21apj}). The end of the activity around 20 Myr ago (see \cref{subsubsec:speclobes}) could have been followed by a jet axis flip by a large angle (nearly $90^{\circ}$). As a consequence, the following (and present) jet activity, started within the last $\sim$10 Myr, is oriented in a different direction. 
\par A second idea involves the superposition of two wobbling modes. Besides precession, the nutation motion of jets has also been predicted in the context of jet dynamics \citep{Gangardt21} and observationally confirmed on pc scales \citep{Britzen18,Dominik21,vonFellenberg23}. We can hypothesize that the jets in RBS~797 experience a long period nutation of the jet axis modulated by a shorter time precession motion. A long period nutation (a few tens of Myr) with a wide nutation angle would explain how the large E-W lobes are orthogonal to the current N-S jets, and a small period precession (a few Myr) would account for the nearly coeval lobes on 50 kpc scales and the S-shaped jets on 10 kpc scales. 
\par Interestingly, all the above processes are usually associated with the presence of binary SMBHs in the central engine (e.g., \citealt{Merritt02,Britzen18}). Below we discuss how the spectral properties of the radio plasma can provide further insights.

\subsubsection{The connection between the inner jets and the outer lobes: injection index and jet power}\label{subsubsec:jetlobeconnection2}
The injection index of the radio plasma has changed between the previous activity that produced the large E-W and N-S radio lobes ($\Gamma\sim0.54$, see \cref{subsubsec:speclobes} and Tab. \ref{tab:lobefit}) and the current activity of the north -- south jets ($\Gamma\sim0.9$, see \cref{subsubsec:specjets}). This indicates that the initial energy distribution of the electrons powering the current north -- south jets was characterized by a relatively higher number of particles at low energies. We noted in \cref{subsubsec:specjets} that the steep injection index of the jets is consistent with those found in powerful FR~IIs. Conversely, the injection index of the radio lobes is consistent with the range of $0.5\leq\Gamma\leq0.7$ expected for radio galaxies (including FR~Is, e.g., \citealt{Laing13}). It has been argued that the injection index positively scales with the jet mechanical power \citep{Blundell99,Konar13}. This scaling would agree with powerful FR~IIs having steeper injection indices. Interestingly, \citet{Konar13} also reported that, in sources that show multiple episodes of activity, a difference in the injection index is expected if the jets change their direction from one outburst to the other. 
\par These arguments are particularly relevant in the case of RBS~797. Specifically, the steeper $\Gamma$ of the youngest AGN activity supports a recent transition from an FR~I radio galaxy (the older, large lobe phase on 50 kpc scales that carved out the E-W and N-S cavities) to a more powerful FR~II (the current north -- south twisted jets on $\sim$5-10 kpc scales). The morphology of the radio emission is also in agreement with this interpretation (see Fig. \ref{fig:jvlalofar}): the diffuse morphology of the large radio lobes and the presence of hotspots in the north -- south jets favours a FR~I interpretation of the former activity and a FR~II interpretation of the latter \citep{Fanaroff74}. 
\par This transition may have been driven by a mechanism that could either be extrinsic or intrinsic to the central engine. In the first case, a change in the accretion rate of the SMBH may have determined an increase of the mechanical power. RBS~797 is a strongly cooling galaxy cluster, and the rate at which gas cools out of the ICM and ultimately feeds the AGN is expected to flicker over time (e.g., \citealt{Gaspari12,Li15}). In the second scenario, the conditions of the central engine may have changed between the outbursts. It has been previously claimed that RBS~797 may host (or may have hosted in the past) binary SMBH \citep{Gitti06,Gitti13,Ubertosi21apj}. If the binary SMBHs merged between the older and the younger outbursts, this may have caused a flip of the new SMBH spin \citep{Merritt02}, which would explain the different direction of the new jet activity. The sudden increase in black hole mass may also be linked to higher -- power jets \citep{Blandford77}. Alternatively, the older and younger outbursts may not have been driven by the same SMBH: if the binary has not coalesced, it is possible that after one SMBH switched off (the one producing the lobe emission) the second SMBH turned on in a different direction. If so, it would be unsurprising to observe different jet powers in successive outbursts.
     \begin{figure*}[ht!]
     \sidecaption
   \centering
   \includegraphics[width=0.7\linewidth]{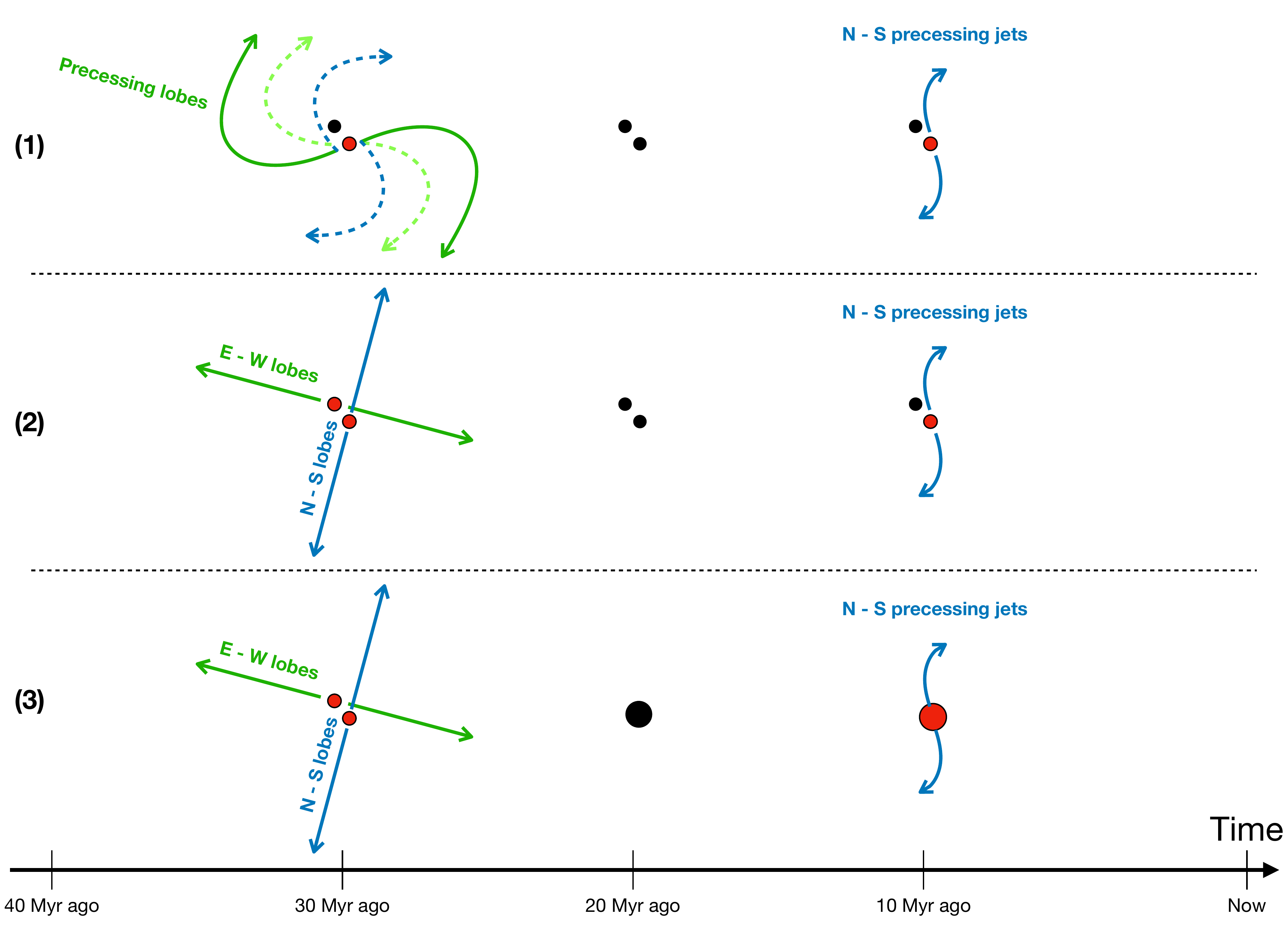}
      \caption{Schematic representation of the three scenarios involving binary SMBHs at the center of the radio galaxy in RBS~797. See \cref{subsec:binary} for details. Arrows represent jets, while circles represent SMBHs (red indicates an active one, black indicates a quiescent one). Blue lines represent the jet activity in the N-S direction, while green lines represent the jet activity in the E-W direction. The dashed lines in panel 1 reflect the precession of the radio lobes on $\sim$50 kpc scales.
              }
         \label{fig:sketch}
   \end{figure*}   
\subsubsection{Alternative scenarios for the outer lobes and inner jets}\label{subsubsec:alternative}
An alternative mechanism that has been invoked to explain secondary lobes of radio galaxies, misaligned from the main jets, is that of backflows. We thus considered the hypothesis that the N-S cavities and corresponding radio lobes resulted from a backflow of the jets in the east-west direction. However, this model is unlikely to be applicable to RBS 797, given the evidence listed as follows.
(1) Backflow formation: plasma from FR II-type jets can be backflowed towards the core owing to pressure imbalance with the surrounding thermal medium. If the environment is highly asymmetric, the backflowed plasma can be diverted towards the direction of the minimum pressure path, usually the minor axis of the host galaxy (e.g., \citealt{leahy1984,capetti2002,saripalli2009}). In RBS 797, the E-W lobes are already aligned with the minor axis of the host galaxy (see the optical images in \citealt{Cavagnolo2011}), and thus it is unlikely that the backflow could traverse the major axis and excavate cavities by displacing the hot gas, as in that direction the gas is much denser. (2) Spectral index and plasma age: backflows typically exhibit steeper spectral indices and longer radiative ages compared to the plasma of the main lobes due to their greater distance from the site of the most recent particle acceleration (e.g., \citealt{mckean2016,cotton2020,morganti2021,Brienza20,brienza2023}). In this respect, the similar age and spectral index of the perpendicular lobes in RBS 797 argue against a backflow model. (3) Radio galaxy classification: backflows are generally associated with FR II radio sources, where they represent plasma escaping the hotspots and returning towards the galaxy (e.g., \citealt{leahy1984,hodges-kluck2011,whitehead2023}). However, the AGN activity responsible for inflating the east-west radio lobes in RBS797 is classified as FR I based on radio power and morphology (this work; \citealt{Gitti06}). 

For the north-south jets on kpc scales, we consider the alternative hypothesis that the jets are bent due to the interaction with the surrounding gas. In this scenario, the four compact bright regions (Fig. \ref{fig:jvlalofar}c) are not hotspots but rather jet knots. However, we find that this is unlikely the case for the jets in RBS 797. High-power jets from BCGs in cool core clusters are hardly bent by the hot and rarified ICM during their earliest stages of propagation (at distances $\lesssim10$~kpc; e.g., \citealt{bourne2023}). On the contrary, such jets typically push aside the hot gas, which has low electron densities, and create cavities. 
In RBS 797, the ICM density within 10 kpc from the center is $\sim$0.1 cm$^{-3}$ \citep{Ubertosi23apj}. This value is typical of strong cool-core clusters (e.g., \citealt{sanderson2009}) and is insufficient to disrupt the jets. While cooler and denser gas phases might in general be more effective in bending a jet, we may likely rule out this possibility as well based on the findings of \citet{Cavagnolo2011} and \citet{calzadilla2022} for RBS 797. 
The BCG of RBS 797 hosts a filamentary nebula of warm gas, a common feature in cool core clusters, which coincides with the region of most efficient cooling in the ICM \citep{Ubertosi23apj}. The nebula is highly asymmetric, being structured in filaments that extend for about 20 kpc only in the south direction. The filaments are spatially correlated with the southern jet \citep{Cavagnolo2011}, but this is more likely explained in the context of jet-induced cooling: jets can induce local instabilities in the ICM, which cools to lower temperatures and produces co-spatial filamentary structures (e.g., \citealt{mcnamara2016}). Thus, we find very unlikely that a dense environment may cause a bending of the jets in RBS 797. In general, and possibly most importantly, we highlight that the observed high degree of symmetry between the north and south jets (in terms of shape, radio brightness, and radio spectral properties) is difficult to reconcile with an external environmental influence, as the environment itself exhibits clear asymmetry. 
\par Ultimately, we caution that the definitions of knots and hotspots are ambiguous. Usually, what separates the two is their location within the radio galaxy: knots are defined as loci along the jet path where the jet encounters material and the plasma is accelerated, whereas hotspots are termination points of the jet flow (e.g., \citealt{Hardcastle07}). However, numerous observations challenge such simple strategy. For instance, the existence of multiple bright compact features associated with the jet termination indicates that jets do not always terminate at a single location \citep{hardcastle2008}. Previous works studying precessing jets labelled as “hotspots” the compact features found along the twisting jet path (e.g., \citealt{cox1991,Hardcastle07,Krause19,Horton20,Horton23}), because in precessing jets the termination point of the ejected plasma changes with time, due to the different launching angle. Given these arguments and considering that jet precession most likely explains the observed properties of the jets in RBS 797, we opted for the term ”hotspots” in this work.

\subsection{Is the central AGN powered by binary SMBHs?}\label{subsec:binary}
Over time, the AGN activity in RBS~797 has been characterized by different processes, including a clear jet precession and possibly a reorientation event between the older and the most recent outbursts. Leveraging the morphological and spectral information on this radio galaxy, we found strong independent evidence for claiming the existence of binary SMBHs in the core of RBS~797. These include: 
\begin{itemize}
    \item The clear twisted morphology of the jets on kpc scale and the existence of multiple, coeval hotspots along these structures. These features are strongly indicative of a precession motion, which in turn is usually attributed to the presence of massive binary SMBHs orbiting each other (e.g., \citealt{Nandi21,Horton23}).
    \item The existence of the perpendicular and equidistant E -- W and N -- S lobes (and corresponding X-ray cavities) on larger scales ($\sim$50 kpc). The dynamical estimates (from X-ray data, \citealt{Ubertosi21apj}) and the radiative estimates (this work, \cref{subsubsec:speclobes}) nicely agree on the fact that both pairs of lobes/cavities were formed around 30 Myr ago and are coeval (with a $\sim$10 Myr uncertainty). The analysis of the radio spectrum further revealed that the jet activities that powered these outbursts had a very similar duration of the active phase ($\sim$10 Myr) and of the fading phase ($\sim$20 Myr). 
\end{itemize}
Additionally, the putative variability of the radio core would be expected from binary SMBHs models (e.g., \citealt{Kun14,Komossa23,Gutierrez23}). However, given the unconfirmed nature of such variability we do not further address this topic.
\par We are aware that interactions between a single SMBH and its accretion disc can also induce jet precession \citep{Baarden75}, due to a forced alignment of the SMBH spin and the disk spin. The precession rate is expected to be slow, and the change in the jet orientation is a single event. This model would be hard to apply to RBS~797: not only we have evidence for multiple changes in jet orientation over time (see also \citealt{Ubertosi21apj,Ubertosi23apj}), but most importantly the Baarden - Petterson effect would not be able to account for the perpendicular, equidistant and coeval outer lobes (see also \citealt{Krause19} for similar arguments). 
\par Therefore, in the following we assume that the presence of binary SMBHs in RBS~797 is the most likely explanation for the morphological and spectral radio properties. Based on this assumption, we can envisage three possible scenarios to explain the outburst history of the central radio galaxy. These are presented below (see also the schematic view of Fig. \ref{fig:sketch}):
\begin{enumerate}
    \item {\it Binary SMBHs with one of them active and an evolving precession (Fig. \ref{fig:sketch}, panel 1)}. If precession was already in place 30 Myr ago with a wide opening angle ($\geq$45$^{\circ}$, to explain the perpendicular position of the outer lobes) and a small precession period ($<$10 Myr, to explain the similar age of the structures), then the formation of a multi-lobed morphology on tens of kpc scales would be expected \citep{Nolting23,Horton23}. Following this scenario (upper row in Fig. \ref{fig:sketch}), after a period of inactivity one of the SMBH in the binary system started again its activity in the N -- S direction (producing the N-S twisted jets on $\sim$10 kpc scales, see Fig. \ref{fig:jvlalofar}), with the gravitational perturbation of the secondary component still causing the jets to precess. Assuming that a single jet episode created both the N -- S and E -- W outer lobes would be consistent with the very similar injection spectral index $\Gamma\sim0.54$ of the perpendicular structures on 50 kpc scales (see \cref{subsubsec:jetlobeconnection2}). The change in jet precession axis from the older (on 50 kpc scales) to the younger (on 10 kpc scales) activity may have been induced either by a sudden change of the SMBH spin axis, or by a long-term nutation motion of the jet direction (see \cref{subsubsec:jetlobeconnection1}).
    \item {\it Binary SMBHs, both of them active 30 Myr ago and only one active now (Fig. \ref{fig:sketch}, panel 2)}. Another possibility to explain the coeval origin of the perpendicular outer lobes is to hypothesize that both SMBHs were active $\sim$30 Myr ago (middle row in Fig. \ref{fig:sketch}; see also \citealt{Ubertosi21apj}). This scenario would not require jet reorientation to occur between the older and the younger outbursts: one of the SMBHs would simply renew its jets in the N -- S direction, and the other would remain silent. 
    \item {\it Binary SMBHs, both of them active 30 Myr ago, then a merger around 10 Myr ago and the resulting single SMBH active now (Fig. \ref{fig:sketch}, panel 3)}. This case assumes a coeval activity around 30 Myr ago of both SMBHs (as the second scenario), but includes the coalescence of the two SMBHs into a single SMBH of higher mass in between the older and younger outbursts (last row in Fig. \ref{fig:sketch}). On the one hand, it would be quite unlikely to serendipitously catch a post-merger SMBH that went through AGN activity before and after the event. On the other hand, this scenario would also explain (contrarily to the first two) other results that we presented. A merger between SMBHs is the key mechanism behind the spin-flip of the central engine \citep{Merritt02}, that is the sudden and large change in jet pointing direction (as observed in RBS~797 between the older and younger outbursts). Additionally, a change in the intrinsic properties of the AGN (in this case, the increased SMBH mass) would be compatible with the observed change of synchrotron injection index \citep{Konar13} and the likely increase of the jet power \citep{Blandford77}, as we discussed in \cref{subsubsec:jetlobeconnection2}.
\end{enumerate}
The above scenarios are all based on indirect evidences of binary SMBHs in RBS~797. 
\par {\it Do we have direct evidence for the presence of two compact radio cores in the available data?} The VLBI observations presented in \cref{subsec:pc} clearly detect a single, compact component at the center of the radio galaxy. If there was a second SMBH on tens of pc scales, it may be undetected owing to a low level of activity. Based on the sensitivity at 1.4 GHz of the VLBA data, we estimate a 3$\sigma$ upper limit on the radio power of any radio-compact component on tens of pc scales of $P_{1.4\text{GHz}}\leq3\times10^{22}$~W/Hz. 
\par Therefore, we tend to exclude that a secondary SMBH is present above tens of pc scales. In this context, it is interesting to note that the precession period caused by the geodetic perturbation of the SMBH spin can be linked to the binary separation. Following \citet{Krause19,Horton20} we can place an upper limit on the binary separation with the following expression:
\begin{equation}
    d_{pc}<0.18\,P^{2/5}_{Myr}\,M^{3/5}_{9}
\end{equation}
where $d_{pc}$ is the binary separation in parsec, $P_{Myr}$ is the precession period in Myr and $M_{9}$ is the active black hole mass in units of $10^{9}$ M$_{\odot}$. The assumptions behind the above equation are that the binary SMBHs are orbiting each other in a circular orbit, and that the observed precessing jets are launched by the more massive black hole. We measured in \cref{subsubsec:modelprec} a precession period of $P = 9.1\pm1.6$ Myr. The mass of the SMBH in RBS~797 has been estimated to be around $1.5\times10^{9}$ M$_{\odot}$ \citep{Cavagnolo10,Ubertosi23apj}. We obtain an upper limit on any binary separation of $d_{pc} < 0.6$ pc. This is an order of magnitude smaller than the highest resolution that can be achieved with the available VLBI data (5 pc $\sim$ 1 mas in the 8.4 GHz VLBA observations). Therefore, the VLBI data do not contradict the binary interpretation.
\par Overall, even if both SMBHs were active (i.e. potentially visible through their radio emission), they would be hard to resolve at the redshift of RBS~797 (z = 0.354). An alternative method to investigate this point would be to search for double-peaked broad and narrow optical lines (e.g., [OIII]) in the core of the BCG, potentially caused by the orbiting binary (see e.g., \citealt{2010ApJ...716..866S}). At present, the existing optical spectral data of the BCG in RBS~797 \citep{fischer1998} are not suited for this type of analysis, and follow-up observations are vital.

%-------------------------------------- Two column figure (place early!)
%   \begin{figure*}
%   \centering
   %%%\includegraphics{empty.eps}
   %%%\includegraphics{empty.eps}
   %%%\includegraphics{empty.eps}
%   \caption{Adiabatic exponent $\Gamma_1$.
%               $\Gamma_1$ is plotted as a function of
%               $\lg$ internal energy $\mathrm{[erg\,g^{-1}]}$ and $\lg$
%               density $\mathrm{[g\,cm^{-3}]}$.}
%              \label{FigGam}%
%    \end{figure*}
%

%--------------------------------------------------- One column table
%   \begin{table}
%       \caption[]{VLBI data presented in this article.}
%          \label{VLBIdata}
%      $$ 
%          \begin{array}{p{0.5\linewidth}l}
%             \hline
%             \noalign{\smallskip}
%             Source      &  T / {[\mathrm{K}]} \\
%             \noalign{\smallskip}
%             \hline
%             \noalign{\smallskip}
%             Yorke 1979, Yorke 1980a & \leq 1700^{\mathrm{a}}     \\
% %           Yorke 1979, Yorke 1980a & \leq 1700             \\
%             Kr\"ugel 1971           & 1700 \leq T \leq 5000 \\
%             Cox \& Stewart 1969     & 5000 \leq             \\
%             \noalign{\smallskip}
%             \hline
%          \end{array}
%      $$ 
%   \end{table}
%

\section{Summary and Conclusions}\label{sec:conclusion}
The multi-scale and multi-frequency radio data presented in this work represent a crucial complement to the deep {\it Chandra} data presented by \citet{Ubertosi21apj,Ubertosi23apj}. While the X-ray observations gave clear insights into the {\it footprints} of feedback on the ICM, the radio observations revealed the direct details of jet evolution in the cluster RBS~797. Here we summarize our results. 
   \begin{enumerate}
      \item The LOFAR (144 MHz, with international stations) and JVLA (3 GHz, 5.5 GHz, 9 GHz) data allowed us to unveil the morphological and spectral properties of the radio galaxy between 10 -- 50 kpc (the scale of the perpendicular pairs of radio lobes) and between 2 -- 10 kpc (the scale of the inner jets). We present the first images of the large scale lobes at sub-arcsecond resolution at 144 MHz (0.3$''$ resolution) and 3 GHz (0.9$''$ resolution). These images confirm the co-spatiality of the radio lobes with the perpendicular X-ray cavities. By fitting a CI-off model to the observed spectra, we find that the radiative age of the E-W lobes ($31.4\pm6.6$~Myr) and of the N-S lobes ($32.1\pm9.9$~Myr) supports a coeval origin of the perpendicular outbursts, that also have similar {\it active phase} duration ($\sim$12~Myr) and {\it passive phase} duration ($\sim$19~Myr). These ages are in good agreement with the X-ray estimates of the cavities and shock ages. Based on the 144 MHz -- 9 GHz emission from the inner N-S jets (on scales of $\leq10$~kpc), we (a) confirm the S-shaped jet morphology; (b) show the presence of two hotspots per jet with a similar radiative age; (c) estimate the age of the twisting N-S jets to be less than $\sim$8~Myr. Based on these results, we determine that jet precession (with period $\sim$9~Myr and jet advance speed $\sim$0.01$c$) can explain the morphological and spectral properties of the N-S twisted jets. We also find that the injection index of the synchrotron emission has steepened between the large scale, older outburst ($\Gamma\sim0.5$) and the younger S-shaped jets ($\Gamma\sim0.9$). This evolution in injection index may suggest that the central radio galaxy has transitioned from an FR~I-like activity to an FR~II-like one around $\sim$10 Myr ago, when the N-S jet activity started.
      \item The e-Merlin observations at 1.6 GHz show the precessing N-S jets at the highest resolution currently available (0.15$''$). With the e-Merlin 5 GHz observations (0.035$''$ resolution), the core emission is resolved into two components (separated by 250 pc), that we interpret as the radio core (component A, $S = 0.55\pm0.04$~mJy) and diffuse, remnant emission from the previous outbursts (component B, $S = 0.38\pm0.04$~mJy). 
      \item The EVN data at 1.6 GHz and 5 GHz reveal the presence of a single, compact core at the heart of RBS~797 (component A). The VLBA data are in agreement with the EVN detection of a single component, but we also find evidence in the 1.6 GHz VLBA data of a faint jet-like feature headed south (component S1, $S = 0.27\pm0.04$~mJy at 1.6 GHz). We also show that there is evidence for monthly-scale time variability of the core radio emission. There is a sususpicious (but possibly serendipitous) systematic flux density difference between the EVN and VLBA observations, which may indicate that the variability is not physical. However, we find that the flux measured with the VLA between 2004 and 2022 has also varied (with a significance of 2.9$\sigma$). Future ad-hoc monitoring is required to confirm the radio core variability. 
      \item Piecing together these results, we discussed that a backflow scenario is not supported, whereas we argued that RBS~797 likely hosts (or hosted) binary active SMBHs in the past. The idea of binary black holes is in agreement with (a) the coeval origin of the large, perpendicular radio lobes (and corresponding X-ray cavities); (b) the very similar active and inactive phase duration of the perpendicular radio lobes; (c) the clear jet precession of the N-S jets; (d) the sudden transition in energy budget, jet power, and synchrotron injection index between the successive outbursts. Radio variability (if confirmed) would be an additional clue. We stress that the detection of a single component in the VLBI data is still consistent with the binary black hole interpretation: the predicted separation of the binary SMBHs (0.6~pc) is an order of magnitude smaller than the resolution of the available radio data (5~pc).
   \end{enumerate}
Our study highlights how a multi-frequency approach can help to discriminate between different scenarios. It also underlines the critical need for high resolution observations that are also sensitive to large scale, diffuse radio emission (such as the LOFAR data with international stations, or the 3 GHz JVLA data). This is required to connect the fine structure of the jets in radio galaxies with the complex geometry of their large expanding lobes. 
\par This work adds up to the handful comparisons between radiative ages and X-ray dynamical timescales of radio lobes and corresponding X-ray cavities (see \citealt{Biava2021} and \citealt{kolokythas2020}). Interestingly, in the case of RBS~797 the radio lobes, the X-ray cavities, and the associated weak shock front have consistent ages down to only a few Myr of uncertainties, thus being a critical test for the AGN -- ICM connection.

\begin{acknowledgements}
The authors thank the referee for their clear suggestions, which improved the scientific quality of this manuscript. F. Ubertosi thanks F. Brighenti for the insightful discussions. e-MERLIN is a National Facility operated by the Univecrsity of Manchester at Jodrell Bank Observatory on behalf of STFC. The European VLBI Network is a joint facility of independent European, African, Asian, and North American radio astronomy institutes. Scientific results from data presented in this publication are derived from the following EVN project code(s): EG080A, EG080B, RSG05, RSG07. The National Radio Astronomy Observatory is a facility of the National Science Foundation operated under cooperative agreement by Associated Universities, Inc. This paper is based (in part) on data obtained with the International LOFAR Telescope (ILT) under project code LC10\_010. LOFAR \citep{vanHaarlem2013} is the Low Frequency Array designed and constructed by ASTRON. It has observing, data processing, and data storage facilities in several countries, that are owned by various parties (each with their own funding sources), and that are collectively operated by the ILT foundation under a joint scientific policy. The ILT resources have benefitted from the following recent major funding sources: CNRS-INSU, Observatoire de Paris and Université d'Orléans, France; BMBF, MIWF-NRW, MPG, Germany; Science Foundation Ireland (SFI), Department of Business, Enterprise and Innovation (DBEI), Ireland; NWO, The Netherlands; The Science and Technology Facilities Council, UK; Ministry of Science and Higher Education, Poland[7]. EDR is supported by the Fondazione ICSC, Spoke 3 Astrophysics and Cosmos Observations. National Recovery and Resilience Plan (Piano Nazionale di Ripresa e Resilienza, PNRR) Project ID CN\_00000013 "Italian Research Center for High-Performance Computing, Big Data and Quantum Computing" funded by MUR Missione 4 Componente 2 Investimento 1.4: Potenziamento strutture di ricerca e creazione di "campioni nazionali di R\&S (M4C2-19)" - Next Generation EU (NGEU). MB acknowledges support from INAF under the Large Grant 2022 funding
scheme (project "MeerKAT and LOFAR Team up: a Unique Radio Window on Galaxy/AGN co-Evolution"). AB acknowledges support from the ERC Stg DRANOEL n. 714245. AI acknowledges  funding from the European Research Council (ERC) under the European Union's Horizon 2020 research and innovation programme (grant agreement No. 833824).
\end{acknowledgements}

\bibliographystyle{aa} % style aa.bst
\bibliography{aanda.bib} % your references Yourfile.bib
% \begin{appendix}
% \onecolumn
% \end{appendix}

\end{document}